\documentclass[12pt,a4paper]{scrartcl}
\pdfoutput=1
\begingroup\expandafter\expandafter\expandafter\endgroup
\expandafter\ifx\csname pdfsuppresswarningpagegroup\endcsname\relax
\else
\fi
\usepackage{amsmath}
\usepackage{amssymb}
\usepackage{dsfont}
\usepackage{xspace}
\usepackage{graphicx}
\usepackage{hepnames}
\usepackage{siunitx}
\usepackage{xfrac}
\usepackage{microtype}
\graphicspath{{figures/}}
\def\beq{\begin{equation}}
\def\eeq{\end{equation}}
\def\beqn{\begin{eqnarray}}
\def\eeqn{\end{eqnarray}}

\newcommand{\Sherpa}{\protect\textsc{Sherpa}\xspace}
\newcommand{\Pythia}{\protect\textsc{Pythia}\xspace}
\newcommand{\Herwig}{\protect\textsc{Herwig}\xspace}
\newcommand{\Vincia}{\protect\textsc{Vincia}\xspace}
\newcommand{\Comix}{\protect\textsc{Comix}\xspace}
\newcommand{\Amegic}{\protect\textsc{Amegic}\xspace}
\newcommand{\BlackHat}{\protect\textsc{BlackHat}\xspace}
\newcommand{\OpenLoops}{\protect\textsc{OpenLoops}\xspace}
\newcommand{\GoSam}{\protect\textsc{GoSam}\xspace}
\newcommand{\NJET}{\protect\textsc{Njet}\xspace}
\newcommand{\CSShower}{\protect\textsc{CSShower}\xspace}
\newcommand{\LO}{\protect\textsc{LO}\xspace}
\newcommand{\LOPS}{\protect\textsc{LoPs}\xspace}
\newcommand{\NLOPS}{\protect\textsc{NloPs}\xspace}
\newcommand{\SMCatNLO}{\protect\textsc{S-Mc@Nlo}\xspace}
\newcommand{\MCatNLO}{\protect\textsc{Mc@Nlo}\xspace}
\newcommand{\MEPSatLO}{\protect\textsc{MePs@Lo}\xspace}
\newcommand{\MEPSatNLO}{\protect\textsc{MePs@Nlo}\xspace}
\newcommand{\MENLOPS}{\protect\textsc{MeNloPs}\xspace}
\newcommand{\NLO}{\protect\textsc{NLO}\xspace}
\newcommand{\PDF}{\protect\textsc{PDF}\xspace}
\newcommand{\HepMC}{\protect\textsc{HepMC}\xspace}
\newcommand{\Rivet}{\protect\textsc{Rivet}\xspace}
\newcommand{\aMCfast}{\protect\textsc{aMCfast}\xspace}
\newcommand{\MCgrid}{\protect\textsc{MCgrid}\xspace}
\newcommand{\APPLgrid}{\protect\textsc{APPLgrid}\xspace}
\newcommand{\FastNLO}{\protect\textsc{FastNLO}\xspace}
\newcommand{\APFELgrid}{\protect\textsc{APFELgrid}\xspace}
\newcommand{\MGMCatNLO}{\protect\textsc{MadGraph5\_aMC@NLO}\xspace}

\newcommand{\diff}{\ensuremath{\text{d}}}
\newcommand{\order}[1]{\mathcal{O}\left(#1\right)}
\newcommand{\dst}{\displaystyle}

\newcommand{\obs}{\langle O\rangle}
\newcommand{\ntrial}{n_\text{trial}}
\newcommand{\Ntrial}{N_\text{trial}}
\newcommand{\muQ}{\mu_Q}
\newcommand{\muQsq}{\mu_Q^2}
\newcommand{\alphaS}{\alpha_\text{s}}
\newcommand{\alphaSt}{\tilde{\alpha}_\text{s}}
\newcommand{\fPDF}[3]{f_{#1}(#2,#3)}
\newcommand{\fPDFt}[3]{\tilde f_{#1}(#2,#3)}
\newcommand{\fPDFs}{f}
\newcommand{\fPDFst}{\tilde f}
\newcommand{\muR}{\mu_R}
\newcommand{\muF}{\mu_F}
\newcommand{\muRsq}{\mu_R^2}
\newcommand{\muFsq}{\mu_F^2}
\newcommand{\muRi}[1]{\mu_{R,#1}}
\newcommand{\muFi}[1]{\mu_{F,#1}}
\newcommand{\muRt}{\tilde\mu_R}
\newcommand{\muFt}{\tilde\mu_F}
\newcommand{\muRit}[1]{\tilde\mu_{R,#1}}
\newcommand{\muFit}[1]{\tilde\mu_{F,#1}}
\newcommand{\muRtsq}{\tilde\mu_R^2}
\newcommand{\muFtsq}{\tilde\mu_F^2}
\newcommand{\muRitsq}[1]{\tilde\mu_{R,#1}^2}
\newcommand{\muFitsq}[1]{\tilde\mu_{F,#1}^2}

\newcommand{\muRcore}{\mu_{R,\text{core}}}
\newcommand{\muRcoret}{\tilde\mu_{R,\text{core}}}
\newcommand{\muRcoresq}{\mu_{R,\text{core}}^2}
\newcommand{\muRcoretsq}{\tilde\mu_{R,\text{core}}^2}
\newcommand{\muRcorei}[1]{\mu_{R,\text{core},#1}}
\newcommand{\muFcore}{\mu_{F,\text{core}}}
\newcommand{\muFcoret}{\tilde\mu_{F,\text{core}}}
\newcommand{\muFcoresq}{\mu_{F,\text{core}}^2}
\newcommand{\muFcoretsq}{\tilde\mu_{F,\text{core}}^2}
\newcommand{\muFcorei}[1]{\mu_{F,\text{core},#1}}

\newcommand{\Qvar}{\ensuremath{Q}}
\newcommand{\Qcut}{\ensuremath{Q_\text{cut}}}
\newcommand{\Qcutsq}{\ensuremath{Q_\text{cut}^2}}

\newcommand{\BME}{\mathrm{B}}
\newcommand{\BMEp}{\mathrm{B}^\prime}
\newcommand{\VIME}{\mathrm{VI}}
\newcommand{\VIMEp}{\mathrm{VI}^\prime}
\newcommand{\KPME}{\mathrm{KP}}
\newcommand{\KPMEp}{\mathrm{KP}^\prime}
\newcommand{\RME}{\mathrm{R}}
\newcommand{\RMEp}{\mathrm{R}^\prime}
\newcommand{\DSME}{\mathrm{D}_S}
\newcommand{\DAME}{\mathrm{D}_A}

\newcommand{\DSMEi}[1]{\mathrm{D}_{S,#1}}
\newcommand{\DAMEi}[1]{\mathrm{D}_{A,#1}}
\newcommand{\DSMEpi}[1]{\mathrm{D}_{S,#1}^\prime}
\newcommand{\DAMEpi}[1]{\mathrm{D}_{A,#1}^\prime}
\newcommand{\BbarME}{\overline{\mathrm{B}}}

\newcommand{\HAME}{\mathrm{H}_A}
\newcommand{\HAMEi}[1]{\mathrm{H}_{A,#1}}

\newcommand{\BMEmerge}{\mathrm{B}^\text{merge}}
\newcommand{\RMEmerge}{\mathrm{R}^\text{merge}}
\newcommand{\BbarMEmerge}{\overline{\mathrm{B}}^\text{merge}}

\newcommand{\HAMEimerge}[1]{\mathrm{H}_{A,#1}^\text{merge}}
\newcommand{\DAMEimerge}[1]{\mathrm{D}_{A,#1}^\text{merge}}
\newcommand{\PS}{\mathrm{PS}}
\newcommand{\PStv}{\mathrm{PS}^\text{vt}}
\newcommand{\PSnlops}{\mathrm{PS}_\text{\NLOPS}}
\newcommand{\PStvnlops}{\mathrm{PS}_\text{\NLOPS}^\text{v}}

\newcommand{\tvar}{\ensuremath{t}}

\newcommand{\tIR}{\ensuremath{t_\text{IR}}}
\newcommand{\cRPS}{k_{\alphaS}}
\newcommand{\cRPSi}[1]{k_{\alphaS,#1}}
\newcommand{\cRtPS}{\tilde{k}_{\alphaS}}
\newcommand{\cFPS}{k_{\scalebox{0.6}{\ensuremath{\fPDFs}}}}
\newcommand{\cFPSi}[1]{k_{\scalebox{0.6}{\ensuremath{\fPDFs}},#1}}
\newcommand{\cFtPS}{\tilde{k}_{\scalebox{0.6}{\ensuremath{\fPDFs}}}}
\newcommand{\KPS}{\mathrm{K}}
\newcommand{\KPSp}{\mathrm{K}^\prime}
\newcommand{\KPSover}{\hat{\mathrm{K}}}
\newcommand{\Pacc}{P_\mathrm{acc}}
\newcommand{\Prej}{P_\mathrm{rej}}
\newcommand{\qacc}{q_\mathrm{acc}}
\newcommand{\qrej}{q_\mathrm{rej}}

\newcommand{\jmul}{j}
\newcommand{\jmax}{{j_\text{max}}}
\newcommand{\jmaxnlo}{{j_\text{max}^\text{\textsc{nlo}}}}

\newcommand{\lR}{l_R}
\newcommand{\lF}{l_F}

\newcommand{\cRp}[1]{c_R^{\,\prime\,(#1)}}
\newcommand{\cRpi}[2]{c_{R,#1}^{\,\prime\,(#2)}}

\newcommand{\cFp}[2]{c_{F,#1}^{\,\prime\,(#2)}}
\newcommand{\cFt}[2]{\tilde c_{F,#1}^{\,(#2)}}
\newcommand{\cFb}[2]{\bar c_{F,#1}^{\,(#2)}}

\usepackage{fancyhdr}
\usepackage{multirow}
\usepackage[a4paper,pdfborder={0 0 0}]{hyperref}
\hypersetup{
  pdfauthor={Enrico Bothmann, Marek Schoenherr, Steffen Schumann}, 
  pdftitle={Reweighting QCD matrix-element and parton-shower calculations}
}
\usepackage[format=hang,labelfont=bf,hypcap=true]{caption}

\numberwithin{equation}{section}

\begin{document}

\title{Reweighting QCD matrix-element and parton-shower calculations}
\author{%
    Enrico Bothmann%
    \thanks{%
        II. Physikalisches Institut, Universit\"at G\"ottingen, Germany,\newline
        \mbox{\href{mailto:enrico.bothmann@phys.uni-goettingen.de}{enrico.bothmann@phys.uni-goettingen.de}},
        \mbox{\href{mailto:steffen.schumann@phys.uni-goettingen.de}{steffen.schumann@phys.uni-goettingen.de}}%
    }
    \and Marek Sch\"onherr
    \thanks{%
        Physik-Institut, Universit\"at Z\"urich, 8057 Z\"urich, Switzerland,
        \href{mailto:marek.schoenherr@physik.uzh.ch}{marek.schoenherr@physik.uzh.ch}%
    }
    \and Steffen Schumann$^*$%
}
\date{}

\fancyhf{} 
\renewcommand{\headrulewidth}{0pt}
\rhead{MCnet-16-22, ZU--TH--21/16}
\maketitle

\begin{abstract}
\noindent
We present the implementation and validation of the techniques used 
to efficiently evaluate parametric and perturbative theoretical 
uncertainties in matrix-element plus parton-shower simulations 
within the \Sherpa event-generator framework.  By tracing the full 
$\alpha_s$ and PDF dependences, including the parton-shower component, 
as well as the fixed-order scale uncertainties, we compute variational 
event weights on-the-fly, thereby greatly reducing the computational 
costs to obtain theoretical-uncertainty estimates.  

\\[0.5em]
Keywords: QCD, NLO, Monte Carlo generators, Matrix Elements, Parton Showers
\end{abstract}

\thispagestyle{fancy}
\tableofcontents

\section{Introduction}

The first operational run of the LHC collider during the 
years 2009-2013 was a tremendous success, clearly culminating 
in the announcement of the discovery of a Higgs-boson candidate
by the ATLAS and CMS collaborations in July 2012 
\cite{Aad:2012tfa,Chatrchyan:2012xdj}. Through a large number
of experimental analyses, focusing on a variety of 
final states and observables, the LHC experiments (re)established 
and underpinned to an unprecedented level of accuracy the validity
of the Standard Model of particle physics (SM)~\cite{Schorner-Sadenius:2015cga}.

When comparing theoretical predictions with actual collider
data, Monte-Carlo event generators prove to be an indispensable
tool. In particular parton-shower Monte-Carlo programmes like  
\Herwig~\cite{Bellm:2015jjp,Bahr:2008pv}, \Pythia~\cite{Sjostrand:2014zea} and 
\Sherpa~\cite{Gleisberg:2003xi,Gleisberg:2008ta} provide 
simulations at the level of exclusive particle-level final 
states~\cite{Buckley:2011ms}. The cornerstones of these generators 
are their implementations of QCD parton-shower algorithms and their 
modelling of the non-perturbative parton-to-hadron fragmentation 
process. With the advent of sophisticated techniques to combine
parton-shower simulations with exact higher-order QCD calculations 
at leading~\cite{Catani:2001cc,Krauss:2002up}, 
next-to-leading~\cite{Frixione:2002ik,Nason:2004rx} 
and even next-to-next-to-leading 
order~\cite{Hamilton:2013fea,Karlberg:2014qua,Hoeche:2014aia,Hoche:2014dla}, 
Monte-Carlo simulations have developed into high-precision tools, 
encapsulating the best of our current knowledge of perturbative QCD. 
 
With these simulations being widely used for making SM predictions, 
e.g.\ of the background expectation in searches for New Physics or the 
detailed properties of Higgs-boson production final states, a
comprehensive and efficient evaluation of associated theoretical
uncertainties is of utmost importance. A comprehensive list of 
sources for generator uncertainties has been quoted in 
\cite{Bellm:2016rhh}.  Following the categories identified 
there, when focusing on systematics related to the perturbative 
phases of event evolution, the following uncertainties might be 
distinguished:  
\begin{itemize}
\item {\em Parametric} uncertainties reflecting the dependence of the 
prediction on input parameters such as couplings, particle masses
or the parton-density functions (PDFs).
\item {\em Perturbative} uncertainties originating from the fact that
perturbation theory is used in making predictions, to fixed-order 
in the matrix elements and resummed to all-orders with a certain 
logarithmic accuracy in the showers, thereby, however, neglecting 
higher-order contributions. Similarly, the use of the large-$N_c$ 
approximation in the showers belongs in this category.
\item {\em Algorithmic} uncertainties corresponding to the actual
choices made in the implementation of the shower algorithm, i.e. 
for the evolution variable, the inclusion of non-singular terms 
in the splitting functions, or the employed matching/merging 
prescription. Per construction, for sensible choices, these systematics 
also correspond to higher-order perturbative corrections, but might
be addressed separately. 
\end{itemize}
In addition to the listed categories, generically non-perturbative effects
such as hadronisation or the underlying event are described through
phenomenological models that feature various generator-specific
choices and parameters, typically subject to tuning against 
experimental data, see for instance~\cite{Buckley:2009bj,Skands:2010ak}. 

This publication focuses on the efficient evaluation of parametric 
and (some) perturbative uncertainties in matrix-element plus parton-shower 
simulations within the \Sherpa event-generator framework. We present
a comprehensive approach to fully trace the $\alpha_s$ and PDF 
dependences in the matrix-element and parton-shower components of 
particle-level \Sherpa simulations in leading- \cite{Hoeche:2009rj} 
and next-to-leading \cite{Hoche:2010kg} order merged calculations
based on the \Sherpa dipole-shower implementation~\cite{Schumann:2007mg}. 
Furthermore, we provide the means to quickly evaluate the renormalisation- 
and factorisation-scale dependence of the fixed-order matrix-element 
contributions. Our approach is based on event-wise reweighting and
allows us to provide with a single generator run a set of variational
event weights corresponding to the predefined parameter and scale
variations, that would otherwise have to be determined through 
dedicated re-evaluations. 
The alternative event weights can either be accessed through the output of a
\HepMC event record~\cite{Dobbs:2001ck}, or directly passed via the internal
interface of \Sherpa to the \Rivet analysis framework~\cite{Buckley:2010ar}.

The systematics of leading-order parton-shower simulations with 
\Herwig7 have recently been discussed in \cite{Bellm:2016rhh}, 
a corresponding reweighting procedure has been presented in 
\cite{Bellm:2016voq}. A similar reweighting implementation for 
the \Pythia 8 parton shower has also appeared recently \cite{Mrenna:2016sih}.
A discussion of uncertainty estimates for the \Vincia shower 
model can be found in \cite{Giele:2011cb,Larkoski:2013yi,Fischer:2016vfv}. A 
comprehensive comparison of various generators is presented in 
\cite{Badger:2016bpw}. The impact of PDFs in parton-shower
simulations has been discussed in \cite{Gieseke:2004tc,Buckley:2016caq}.

Our paper is organised as follows. In Sec.~\ref{sec:fo} we review the
dependence structure of leading-order (LO) and next-to-leading-order
QCD calculations on $\alpha_s$, the PDFs and the renormalisation 
and factorisation scales, and introduce the reweighting approach. 
In Sec.~\ref{sec:parton_shower} we extend this to parton-shower
simulations and in particular the algorithm employed in the 
\Sherpa framework. In Sec.~\ref{sec:multijet_merging} we present
the generalisation of the reweighting approach to multijet-merged
calculations, based on leading and next-to-leading-order 
matrix elements matched to the parton shower. Our conclusions
are summarised in Sec.~\ref{sec:conclusions}.
In App.~\ref{app:timing} we present CPU time measurements that assess the
reduction in computational time when the reweighting is used.  The technical
details on enabling and accessing the variations considered in \Sherpa runs are
listed in App.~\ref{app:keywords}.

Note, while fixed-order reweighting is already available with 
\Sherpa-2.2, the general reweighting implementation described
here, including parton showers and multijet merging, will be part of 
the next release, i.e.\ \Sherpa-2.3.

\section{Reweighting fixed-order calculations}
\label{sec:fo}

In order to re-evaluate a QCD cross-section calculation for a new 
choice of input parameters, i.e.\ $\alphaS$, PDFs or renormalisation
and factorisation scales, it is necessary to understand 
and trace-out its respective dependences.
This is a rather easy task at leading-order (\LO) but is already
more involved when considering next-to-leading order (\NLO)
calculations in a given subtraction scheme. However, these 
decompositions have been presented for Catani--Seymour dipole 
subtraction and the FKS subtraction formalism in~\cite{Bern:2013zja,Frederix:2011ss}. 

In this section, we briefly review the dependence structure
and discuss the corresponding reweighting equations for \LO
and Catani--Seymour subtracted \NLO calculations within 
the \Sherpa framework. With this paragraph we also introduce 
the notation used in the later sections, which explore the 
reweighting of more intricate QCD calculations, involving 
QCD parton showers and merging different final-state multiplicity
processes.

\subsection{The leading-order case}
\label{sec:fo:lo}
A \LO parton-level calculation of some observable or measurement function 
of the final-state momenta $O$ is based on Born matrix elements 
$\BME$ of $\order{\alphaS^n}$. It exhibits explicit dependences on the PDFs 
$f=f_a(x,\muFsq)$, the running strong coupling $\alphaS=\alphaS(\muRsq)$,
the renormalisation scale $\muR$ and the factorisation scale $\muF$:
\begin{equation}
    \obs^\text{\LO}
    =\int\diff\Phi_B\;\BME(\Phi_B)\;O(\Phi_B)
    =\lim\limits_{N\to\infty}\,\frac{1}{\Ntrial}\;
          \sum\limits_{i=1}^N \BME(\Phi_{B,i})\;O(\Phi_{B,i})
    \label{eq:LO}
\end{equation}
with $\Ntrial=\sum_{i=1}^N\left.\ntrial\right._{\!,i}$, $\ntrial$ denoting
the number of attempts to generate an accepted event configuration, and
\begin{equation}\label{eq:defBMEp}
    \BME(\Phi_B) \equiv \BME(\Phi_B;\alphaS,\fPDFs;\muR,\muF)
    = \alphaS^n(\muRsq)\;\fPDF{a}{x_a}{\muFsq}\;\fPDF{b}{x_b}{\muFsq}\;\BMEp(\Phi_B)\,.
\end{equation}
Therein, the $\BME$ contains all couplings, symmetry and flux factors, 
and PDFs, whereas $\BMEp$ has the PDFs, here for assumed 
two incoming parton flavours $a$ and $b$, and the strong coupling stripped off.
Note that we have suppressed the event index $i$ here. It is understood
that $\BME$ depends on the event kinematics and that $\muR$ and
$\muF$ can be chosen dynamically, i.e. in a momentum (and flavour) dependent way. 
Changing the input parameters $\mu_R\to\muRt$, $\muF\to\muFt$, and the 
input functions $\fPDFs\to\fPDFst$, $\alphaS\to\alphaSt$ results in
\begin{equation}\label{eq:LO-scale-var}
  \begin{split}
    \BME(\Phi_B;\alphaSt,\fPDFst;\muRt,\muFt)
    \,=&\;\alphaSt^n(\muRtsq)\;\fPDFst_a(x_a,\muFtsq)\;\fPDFst_b(x_b,\muFtsq)\;
          \BMEp(\Phi_B)\,.
  \end{split}
\end{equation}
From eq.~\eqref{eq:LO-scale-var} we conclude that for PDF reweighting it is necessary
to know the $x_{a,b}$ values of the event.

For an unweighted event generation, the event weights are uniform initially, i.e.\ 
$\BME(\Phi_B;\alphaS,f;\muR,\muF)=w_\text{norm}$, eq.\ \eqref{eq:LO} thus
simplifies to 
\begin{equation}
    \obs^\text{\LO}
    =\lim\limits_{N\to\infty}\,\frac{w_\text{norm}}{\Ntrial}\;
          \sum\limits_{i=1}^N O(\Phi_{B,i})\;.
    \label{eq:LO_unw}
\end{equation}
Scale and parameter variations then work the very same way as for weighted 
events. Applying eq.\ \eqref{eq:LO-scale-var} then, however, leads to a 
broader weight distribution and eq.\ \eqref{eq:LO} has to be used again. 
Partially unweighted events can be treated on the same footing. These 
conclusions hold irrespective of the type of event generation whenever 
(partially) unweighted event generation is possible, i.e.\ when the weight 
distribution is bounded from above and below. We therefore will not comment 
further on it.

\subsection{The next-to-leading-order case}
\label{sec:fo_nlo}

A full NLO parton-level calculation including real-emission and
one-loop corrections of $\order{\alphaS^{n+1}}$ based in 
Catani--Seymour dipole subtraction \cite{Catani:1996vz,Catani:2002hc} 
has the following structure
\begin{equation}
    \label{eq:nlo}
    \begin{split}
        \obs^\text{\NLO} =& \int \diff \Phi_B\, \left[\vphantom{\sum_j}
            \BME(\Phi_B) + \VIME(\Phi_B)
            + \int \diff x_{a/b}^\prime\, \KPME(\Phi_B,x_{a/b}^\prime)
        \right]\,O(\Phi_B)\\
    &{}
        + \int\diff\Phi_R\;\left[\vphantom{\sum_j}
            \RME(\Phi_R)\,O(\Phi_R)
            - \sum\limits_{j}\DSMEi{j}(\Phi_{B,j}\cdot\Phi_1^j)\,O(\Phi_{B,j})
        \right]
    \\
    =& \lim_{N\to\infty}\frac{1}{\Ntrial}
    \left\{
        \sum\limits_{i=1}^{N_B}\left[\vphantom{\sum_j}
            \BME(\Phi_{B,i}) + \VIME(\Phi_{B,i}) + \KPME(\Phi_{B,i},x_{a/b}^\prime)
        \right]\,O(\Phi_{B,i})
    \right.\\
    &\left.{}
          + \sum\limits_{i=1}^{N_R} \left[
          \RME(\Phi_{R,i})\,O(\Phi_{R,i})
          -\sum\limits_{j}\DSMEi{j}(\Phi_{B,j,i}\cdot\Phi_{1,i}^j)\,O(\Phi_{B,j,i})
        \right]
   \right\}\,,\hspace*{-15mm}
  \end{split}
\end{equation}
where the new parts have the following dependences
\begin{equation}
  \begin{split}
    \VIME(\Phi_B) 
    &\equiv \VIME(\Phi_B;\alphaS,\fPDFs;\muR,\muF)\,,\\
    \KPME(\Phi_B,x_{a/b}^\prime) 
    &\equiv \KPME(\Phi_B,x_{a/b}^\prime;\alphaS,\fPDFs;\muR,\muF)\,,\\
    \RME(\Phi_R) 
    &\equiv \RME(\Phi_R;\alphaS,\fPDFs;\muR,\muF)\,,\\
    \DSMEi{j}(\Phi_{B,j}\cdot\Phi_1^j) 
    &\equiv \DSMEi{j}(\Phi_{B,j}\cdot\Phi_1^j;\alphaS,\fPDFs;\mu_{R,j},\mu_{F,j})\,.
  \end{split}
\end{equation}
Therein, $\VIME$ combines the renormalised one-loop matrix element 
with the $\mathrm{I}$-operator of the Catani-Seymour subtraction scheme. This
operator gives the flavour-diagonal endpoint contribution of the integrated
subtraction terms. $\VIME$ is thus separately infrared finite and exhibits a
common transformation behaviour. Thus, for 
$\alpha\to\alphaSt$, $\fPDFs\to\fPDFst$, $\muR\to\muRt$ and $\muF\to\muFt$ 
\begin{equation}\label{eq:NLO-scale-var}
  \begin{split}
    \lefteqn{\hspace*{-5mm}\VIME(\Phi_B;\alphaSt,\fPDFst;\muRt,\muFt)}\\
    \,=\;&\alphaSt^{n+1}(\muRtsq)\;\fPDFst_a(x_a,\muFtsq)\;\fPDFst_b(x_b,\muFtsq)\;
    \bigg[
    \VIMEp(\Phi_B)
          +\cRp{0}\lR+\tfrac{1}{2}\,\cRp{1}\lR^2
    \bigg]\;,
  \end{split}
\end{equation}
with $\alphaS$- and PDF-independent coefficients 
$\cRp{i}$ and $\lR=\log(\muRtsq/\muRsq)$. Again, $\VIMEp$ is stripped 
of all coupling and PDF factors.

The $\KPME$-terms are defined as the remainders of the integrated 
dipole subtraction terms, containing all flavour changing and 
$x_{a/b}^\prime$-dependent pieces, combined with the collinear 
counterterms. Here, $x_{a/b}^\prime$ are the ratios of the partonic 
momentum fractions in the respective dipole before and after radiation. 
Again, this combination is separately infrared finite and transforms 
as one unit. When evaluated for the modified set of input parameters, 
they read 
\begin{equation}\label{eq:KP-scale-var}
  \begin{split}
    \lefteqn{\hspace*{-5mm}\KPME(\Phi_B,x_{a/b}^\prime;\alphaSt,\fPDFst;\muRt,\muFt)}\\
    \,=\;&\alphaSt^{n+1}(\muRtsq)\,\fPDFst_a(x_a,\muFtsq)\;\fPDFst_b(x_b,\muFtsq)\;
          \KPMEp(\Phi_B,x_{a/b}^\prime;\fPDFst;\muFt)\\
    \,=\;&\alphaSt^{n+1}(\muRtsq)\,
    \bigg[
      \left(
        \fPDFst_a^q \cFp{a}{0} + \fPDFst_a^q(x_a^\prime)\, \cFp{a}{1}
        +\fPDFst_a^g \cFp{a}{2} + \fPDFst_a^g(x_a^\prime)\, \cFp{a}{3}
      \right)
      \fPDFst_b(x_b,\muFtsq)
    \\
         &\hspace*{20mm}{}+
      \fPDFst_a(x_a,\muFtsq)
      \left(
        \fPDFst_b^q \cFp{b}{0} + \fPDFst_b^q(x_b^\prime)\, \cFp{b}{1}
        +\fPDFst_b^g \cFp{b}{2} + \fPDFst_b^g(x_b^\prime)\, \cFp{b}{3}
      \right)
    \bigg]
  \end{split}
\end{equation}
with the coefficients $\cFp{a/b}{i}=\cFt{a/b}{i}+\cFb{a/b}{i}\,\lF$ for 
$i\in\{0,\ldots,3\}$, $\lF=\log(\muFtsq/\muFsq)$, and
\begin{align*}
    \fPDFst_q^q
    \,=\;&\fPDFst_q(x_q,\muFtsq)\,,
    \hspace*{20mm}\vphantom{\sum_q}&
    \fPDFst_g^q
    \,=\;&\sum_q \fPDFst_q(x_g,\muFtsq)\,,\\
    \fPDFst_q^q(x_q^\prime)
    \,=\;&x_q^\prime \fPDFst_q(\tfrac{x_q}{x_q^\prime},\muFtsq)
    \vphantom{\sum_q}\,,&
    \fPDFst_g^q(x_g^\prime)
    \,=\;&x_g^\prime \sum_q \fPDFst_q(\tfrac{x_g}{x_g^\prime},\muFtsq)\,,\\
    \fPDFst_q^g
    \,=\;&\fPDFst_g(x_q,\muFtsq)
    \vphantom{\sum_q}\,,&
    \fPDFst_g^g
    \,=\;&\fPDFst_g(x_g,\muFtsq)\,,\\
    \fPDFst_q^g(x_q^\prime)
    \,=\;&x_q^\prime \fPDFst_g(\tfrac{x_q}{x_q^\prime},\muFtsq)
    \vphantom{\sum_q}\,,&
    \fPDFst_g^g(x_g^\prime)
    \,=\;&x_g^\prime \fPDFst_g(\tfrac{x_g}{x_g^\prime},\muFtsq)\,,
\end{align*}
for $a,b= \{q,\,g\}$, respectively. Thereby, the sum over $q$ includes 
all light-quark flavours, corresponding to all potential quarks emitting 
a gluon.
We note that in order to obtain the reweighted expressions for the $\VIME$ and $\KPME$
contributions the additional book-keeping of the $\cRp{i}$, $\cFt{a/b}{i}$ and 
$\cFb{a/b}{i}$ (altogether 18)\footnote{
  The two parameters $\cRp{i}$ correspond to the single and double pole 
  coefficients of the loop matrix element while the remaining sixteen 
  coefficients are comprised of eight pairs of coefficients, $\cFb{a/b}{i}$ 
  and $\cFt{a/b}{i}$, corresponding to the $\muF$-dependent and -independent 
  parts for all four flavour structures of each beam, respectively.
} coefficients is required~\cite{Bern:2013zja}. 
Due to its composite structure, the $\KPME$-terms do not possess a coupling- 
and PDF-stripped version $\KPMEp$. Nonetheless, we formally introduce 
a still PDF-dependent version $\KPMEp$ in eq.\ \eqref{eq:KP-scale-var} 
for reference in later sections.

The remaining pieces of eq.\ \eqref{eq:nlo} are the Born matrix element 
$\BME$, the real emission contribution $\RME$ and the differential 
dipole subtraction terms $\DSMEi{j}$. The latter defines an underlying 
Born configuration $\Phi_{B,j}$ through its dipole-dependent phase-space 
map, employing the phase-space factorisation $\Phi_R=\Phi_{B,j}\cdot\Phi_1^j$. 
While the transformation of $\BME$ under the exchange of input parameters 
was detailed in eq.\ \eqref{eq:LO-scale-var}, the transformation of $\RME$ and 
the $\DSMEi{j}$ contributions works identically, merely having to adjust the 
power of the strong-coupling factor.

\subsection{Validation}
\label{sec:nlo-validation}
The reweighting approach outlined above has been implemented in the \Sherpa
framework for the two matrix-element generators \Amegic \cite{Krauss:2001iv} 
and \Comix \cite{Gleisberg:2008fv,Hoche:2014kca} in conjunction with the 
corresponding Catani--Seymour dipole-subtraction implementation 
\cite{Gleisberg:2007md}. The required decomposition of virtual amplitudes
is generic and can be used for matrix elements from 
\BlackHat \cite{Berger:2008sj,Bern:2013zja}, 
\OpenLoops \cite{Cascioli:2011va}, \GoSam \cite{Cullen:2014yla}, 
\NJET \cite{Badger:2012pg}, the internal library of simple $2\to 2$ processes,
or, via the BLHA interface \cite{Binoth:2010xt}. 

\begin{figure}[p]
  \centering
  \includegraphics[width=0.9\textwidth]{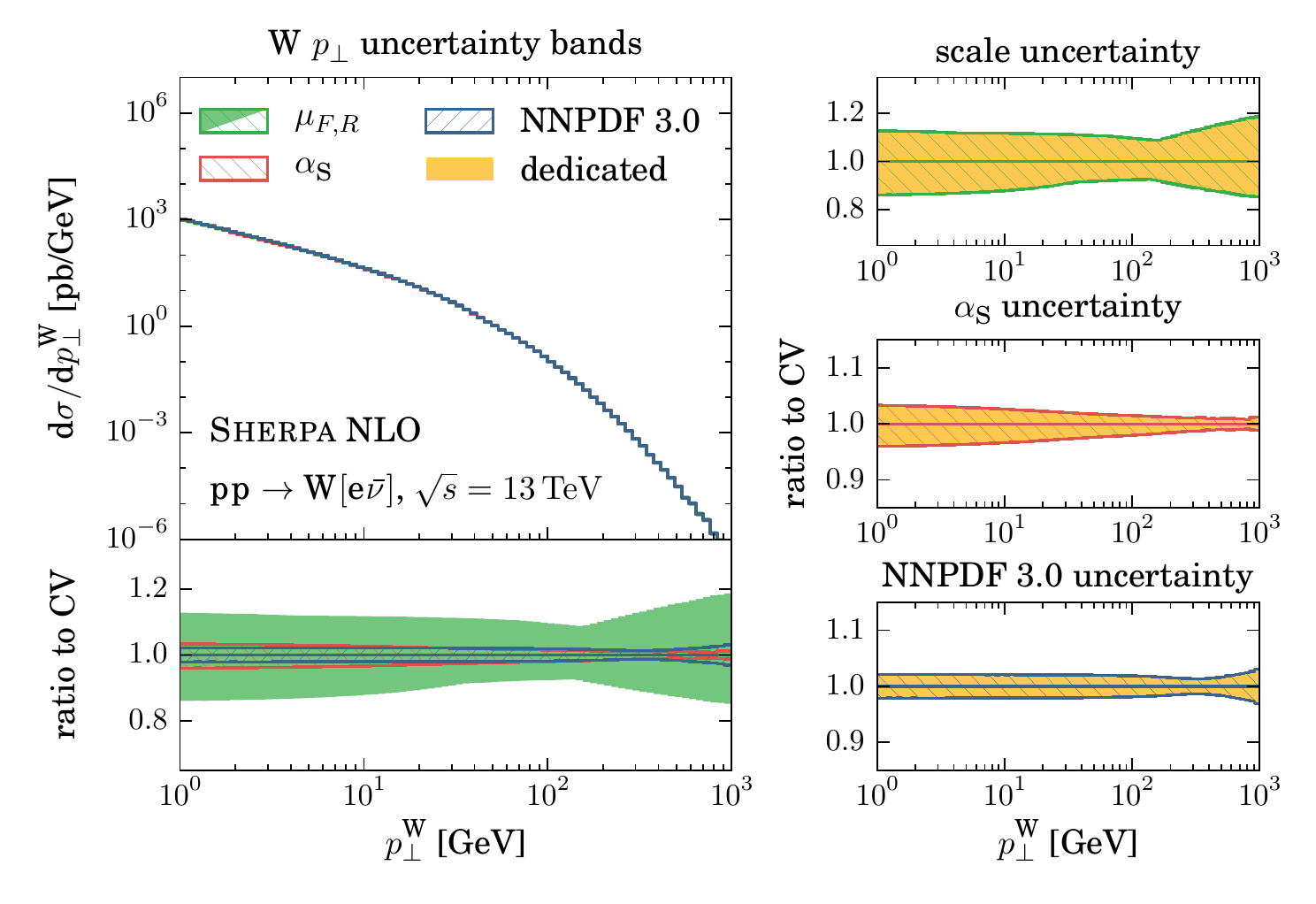}
  \includegraphics[width=0.9\textwidth]{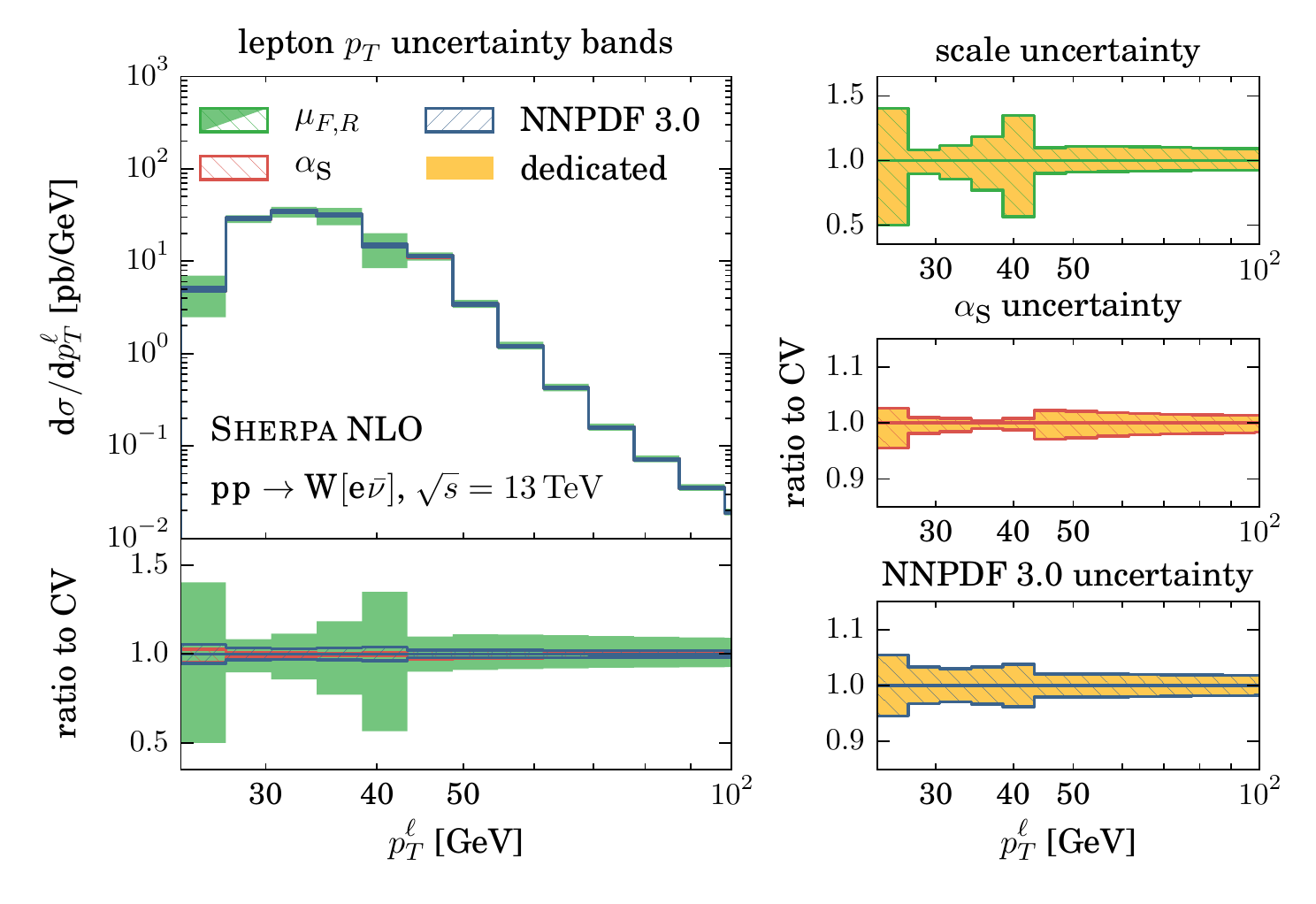}
  \caption{The gauge-boson and lepton transverse momenta in off-shell \PW
      production at the LHC with independent variations of $\mu_{F,R}$ (green),
      $\alphaS$ (red) and the PDF (blue). In the right-hand panels, the
      individual uncertainty bands, calculated via an on-the-fly reweighting,
      are compared to uncertainty bands from dedicated calculations (yellow).
      They are found to be equal.}
  \label{fig:nlo-w-pt}
\end{figure}

Here we shall present the validation of the reweighting approach in particular
of NLO QCD event samples. For that purpose we consider \PW-boson production
in \SI{13}{\TeV} proton-proton collisions at NLO QCD, and focus on the 
transverse-momentum distribution for the \PW and the lepton it decays to.
In Fig.~\ref{fig:nlo-w-pt}, the scale, $\alphaS$ and PDF uncertainty bands for
the \mbox{\PW $p_\perp$} and the lepton $p_\perp$
distributions are presented. All three bands have been produced for both observables using the
internal reweighting of \Sherpa from a single event generation run using $\muF
= \muR = H_T'$ with
\begin{equation}
  H_T' \equiv m_{\perp}^{\Pe\Pnu} + \sum_{j} p_\perp^j\,,
\end{equation}
a scale choice that has been motivated in~\cite{Berger:2010zx}. For the
PDFs the NNPDF 3.0 \NLO set~\cite{Ball:2014uwa} has been used with
$\alphaS(m_{\PZ}^2) = 0.118$. The running of $\alphaS(\muRsq)$ is calculated within
\Sherpa using its renormalisation group equation at NLO with parton thresholds 
as given by the PDF. 

The treatment of partonic thresholds deserves a short discussion. While any 
flavour thresholds in the running of $\alpha_s$ do not present any challenges 
to the reweighting algorithm as $\alpha_s(\mu^2)>0$ for all $\mu^2>0$ and 
any loop order, this is different for the PDFs, where crossing a parton 
threshold results in a vanishing PDF for that flavour. Hence, the cross 
section component of the given partonic channel may be zero if no other 
non-zero contribution exists. Such an event will be discarded and, 
thus, cannot be reweighted. If now the respective parton threshold of the 
target PDF is smaller than the target factorisation scale while the one of 
the nominal PDF is larger than the nominal factorisation scale we are in a 
region of phase space where the reweighting must fail to reproduce 
a dedicated calculation. This could be remedied by storing events as well 
which vanish solely due to crossing PDF thresholds. However, as only observables 
sensitive to on-threshold production of light quarks (typically bottom 
quarks) are susceptible to these effects, they are of little relevance to 
the vast majority of LHC observables.%
\footnote{ A typical example for the threshold problem in the reweighting 
          would be the very low-$p_\perp$ part of $b$-jet spectrum in 
          $Wb$ production in a calculation with five massless flavours. 
          Any strong dependence on the bottom quark PDF threshold, however, 
          also indicates the invalidity of a calculation with five massless 
          flavour for this observable.}

For the {\em scale uncertainty band} we employ a 7-point scale variation for
$\muR$ and $\muF$: Both scales are varied independently by factors of
\sfrac{1}{2} and 2, omitting the variations with ratios of 4 between the two
scales. The uncertainty is then taken as the envelope of all variations.  The
{\em $\alphaS$ uncertainty band} is generated by varying the numerical value of
the starting point of the running coupling, $\alphaS(m_{\PZ}^2)$, to the
following five values: $0.115$, $0.117$, $0.118$, $0.119$ and $0.121$.
Note that this variation of $\alphaS$ should also enter the PDF fit, and hence
the PDFs are varied consistently. This is expected to extenuate the effect of
the $\alphaS$ variation in most cases, as the PDF of the varied $\alphaS$
is still fitted to describe the same data as the PDF of the nominal $\alphaS$.
This consistent $\alphaS$+PDF variation is also part of the PDF4LHC
recommendations for LHC Run II~\cite{Butterworth:2015oua}.  The envelope of
these $\alphaS$+PDF variations is taken as the respective uncertainty.  The
pure {\em PDF uncertainty} estimate is generated using the average and the
standard deviation over the $100$ PDF replicas provided by the NNPDF3.0 set (at
a fixed value of $\alphaS=0.118$). This corresponds to the \SI{68}{\percent}
confidence level.  This set-up is repeated for later reference in
Table~\ref{tab:variations}, along with a CT14 PDF variant, which is used in
later studies.

\begin{table}
    \centering
    \begin{tabular}{@{}|l|c|c|c|@{}}
        \hline
        \vphantom{$\dst \frac{1}{2}$}
        & nominal & variations & error band \\
        \hline \hline
        \vphantom{$\dst \frac{1}{2}$}
        \multirow{2}{*}{PDF sets} & CT14 &
        56 Hessian error sets with a \SI{90}{\percent} CL &
        Hessian \\
        & NNPDF3.0 & 100 statistical replicas with a \SI{68}{\percent} CL& statistical
        \\[0.2em] \hline
        \vphantom{$\dst \frac{1}{2}$}
        $\alphaS(m_Z^2)$ value & 0.118 & 0.115, 0.117, 0.119, 0.121 & envelope
        \\ \hline
        \vphantom{$\dst \frac{1}{2}$}
        $\muR$/$\muF$ factors & $\left(\vphantom{\frac{1}{2}}1, 1\right)$ &
        $\left(\frac{1}{2}, \frac{1}{2}\right)$, $\left(1, \frac{1}{2}\right)$,
        $\left(\frac{1}{2}, 1\right)$, $\left(\vphantom{\frac{1}{2}}2, 1\right)$,
        $\left(\vphantom{\frac{1}{2}}1, 2\right)$, $\left(\vphantom{\frac{1}{2}}2, 2\right)$ &
        envelope
        \\ \hline
    \end{tabular}\vspace*{1ex}
    \vspace*{2mm}
    \caption{Variations, which are used for studies in
        this publication, with two variants depending on the PDF choice.
        Note that each $\alphaS(m_Z^2)$ value is used with its
        associated PDF set variant in the context of hadronic collisions. \label{tab:variations}}
\end{table}

Comparing the uncertainties for the \PW $p_\perp$, we observe that the scale
uncertainties are the
largest, with relative deviations of $\order{\SI{10}{\percent}}$. The relative
deviations related to the PDF and the strong coupling do not exceed
$\sim\SI{3}{\percent}$. The scale uncertainty exhibits a minimum for
$\SI{100}{\GeV} < p_\perp^{\PW} < \SI{200}{\GeV}$.
The reason is that the variations of $\muF$ alone cross the central value
prediction in this range, such that only the $\muR$ variation contributes
to the overall scale uncertainty here.

Note that $p_\perp^{\PW} = 0$ at $\order{\alphaS^0}$, and therefore only
real-emission events contribute to the distribution. Hence, the observable is
only described to leading-order. We introduce it here as a reference for our
later validations including the parton-shower, which use this observable.  For
the current validation, we complement the discussion of the \PW transverse
momentum with the one of the lepton it decays to, as the region below
$m_{\PW}/2$ is already filled at $\order{\alphaS^0}$, and therefore we have in
part a true next-to-leading description for this observable. In fact,
the scale uncertainties are much larger in that region, especially towards the
$m_{\PW}/2$ threshold, and at the lepton $p_\perp$ cut at \SI{25}{\GeV}. This
gives a more realistic picture of the perturbative uncertainties than in the
leading-order region above the threshold.

The small panels on the right of Fig.~\ref{fig:nlo-w-pt} compare
the uncertainty bands calculated using the reweighting approach to 
uncertainty bands where dedicated calculations have been done for each variation.
We observe that all bands overlap perfectly for both observables. This is
because the reweighting
as presented above is exact and for all runs the same phase-space points 
could be used: The reweighted and the dedicated predictions for each variation 
are therefore equal, and so are the uncertainty bands.%
\footnote{ The reweighted and the dedicated calculations are implemented
independently, such that their predictions can vary within the numerical
uncertainties of the calculation. However, these lie several orders of 
magnitude below the physical uncertainties considered here.}


\section{Reweighting parton-shower calculations}
\label{sec:parton_shower}

If parton-showering is added to a \LO calculation, the value of the observable
is not evaluated at $\Phi_B$ any longer, but at $\PS(\Phi_B)$,
which denotes the phase-space point after showering.
Applying this modification to eq.~\eqref{eq:LO} yields
\begin{equation}
    \obs^\text{\LOPS}
    =\int\diff\Phi_B\;\BME(\Phi_B)
    \;\PS(O,\Phi_B)
    =\lim\limits_{N\to\infty}\,\frac{1}{\Ntrial}\;
    \sum\limits_{i=1}^N \BME(\Phi_{B,i})
    \;\PS(O,\Phi_{B,i})\,.
    \label{eq:LOPS}
\end{equation}
Therefore the reweighting for $\BME$ does not need to be altered,
but the parton-shower emissions depend on the PDF, the strong coupling, 
their respective scale prefactors $\cRPS$ and $\cFPS$ (detailed below) 
and the starting scale $\muQ$, i.e.\ 
\begin{equation}\label{eq:PSdep}
  \PS(O,\Phi_B) 
  \equiv \PS(O,\Phi_B;\muQsq)
  \equiv \PS(O,\Phi_B;\cRPS,\cFPS;\alphaS,\fPDFs;\muQsq)\;.
\end{equation}
In order to reweight the parton-shower emissions, we first need to identify
its exact dependence structure.
Schematically, it acts on the phase-space element in the following way
\begin{equation}\label{eq:PSaction}
    \PS(O,\Phi_n;\tvar^\prime)=\;\Delta_n(\tIR,\tvar^\prime)\,
          O(\Phi_n)\vphantom{\int_{\tvar}^{\tvar}}
         +\int_{\tIR}^{\tvar^\prime}
	  \diff\Phi_1\;
	  \KPS_n(\Phi_1)\,
	  \Delta_n(\tvar,\tvar^\prime)\,
	  \PS(O,\Phi_{n+1};\tvar)\,,
\end{equation}
where the Sudakov form factor of the $n$-parton state, $\Delta_n$, and 
its splitting kernel $\KPS_n$ have been introduced. While the 
first term describes the no-emission probability between the starting scale 
$\tvar^\prime$ and the infrared cut-off $\tIR$ and therefore does not 
change the phase-space element, the second term describes the emission 
of a parton at scale $\tvar$ in the configuration 
$\diff\Phi_1=\diff\tvar\,\diff z\,\diff\phi\;J(\tvar,z)$ 
(the integration boundaries are to be understood in this decomposition), 
leading to a configuration $\diff\Phi_{n+1}=\diff\Phi_n\cdot\diff\Phi_1$. 
The Jacobian $J$ is not relevant to the discussion here and is subsequently 
absorbed in the splitting kernel $\KPS_n$. As the emissions 
are ordered in $\tvar$, the Sudakov form factor in the second term ensures 
that the current emission is the hardest after starting the evolution at 
$\tvar^\prime$. 
Additional emissions may occur at smaller $\tvar$ and are not resolved at
this stage -- they are described by the parton shower acting on the newly 
produced state $\Phi_{n+1}$ with the new starting scale $\tvar$. In 
eq.\ \eqref{eq:PSaction} the dependences on $\alphaS$, the PDFs, 
and their respective scale prefactors $\cRPS$ and $\cFPS$ have been omitted 
for brevity. They directly carry over to the splitting kernel and 
the Sudakov form factor, according to 
\begin{equation}
  \begin{split}
    \Delta_n(\tvar_2,\tvar_1;\cRPS,\cFPS;\alphaS,\fPDFs)
    \,=\;\exp\left(-\int_{\tvar_2}^{\tvar_1}
	            \diff\Phi_1\;
	            \KPS_n(\Phi_1;\cRPS,\cFPS;\alphaS,\fPDFs)
	     \right)\,.
  \end{split}
\end{equation}
When considering parton-shower emissions off NLO QCD matrix elements 
special emphasis has to be given to the first emission as described in
Sec.~\ref{sec:nlops} below. 

\subsection{Parton-shower dependence structure}

\begin{table}
    \centering
  \begin{tabular}{|c|c|c|c|c|c|}
  \hline
  \vphantom{$\dst\int$}Type &
    \hspace*{3mm}$\dst z$\hspace*{3mm} & 
    \hspace*{7mm}$\dst y$\hspace*{7mm} & 
    \hspace*{3mm}$\dst x$\hspace*{3mm} & 
    \hspace*{1mm}$\dst (ij,k)\to(i,j,k)$\hspace*{1mm} & 
    \hspace*{1mm}$\dst c,c'$\hspace*{1mm} \\
  \hline\hline
  FF\vphantom{$\dst\int\limits_A^B$} & 
    $\dst \tilde{z}_i$ &
    $\dst y_{ij,k}$ &
    $\dst 1$ &
    $(ij,k)\to(i,j,k)$ &
    $\dst a,a$ \\
  \hline
  FI\vphantom{$\dst\int\limits_A^B$} & 
    $\dst\tilde{z}_i$ &
    $\dst\frac{1-x_{ij,a}}{x_{ij,a}}$ & 
    $\dst x_{ij,a}$ &
    $(ij,a)\to(i,j,a)$ &
    $\dst a,a$ \\
  \hline\hline
  IF\vphantom{$\dst\int\limits_A^B$} & 
    $\dst x_{jk,a}$ &
    $\dst\frac{u_j}{x_{jk,a}}$ &
    $\dst x_{jk,a}$ &
    $(aj,k)\to(a,j,k)$ &
    $\dst aj,a$ \\
  \hline
  II\vphantom{$\dst\int\limits_A^B$} & 
    $\dst x_{j,ab}$ &
    $\dst\frac{\tilde{v}_j}{x_{j,ab}}$ &
    $\dst x_{j,ab}$ &
    $(aj,b)\to(a,j,b)$ &
    $\dst aj,a$ \\
  \hline
  \end{tabular}\vspace*{1ex}
  \vspace*{2mm}
  \caption{%
    Definition of the evolution and splitting variables for each dipole type. 
    The fifth column lists the splitting process as seen from the Born 
    process, $c$ and $c'$ refer to the flavour of the initial state before 
    and after the splitting process, respectively. The variables $y_{ij,k}$, 
    $\tilde{z}_i$, $x_{ij,a}$, $x_{jk,a}$, $x_{j,ab}$, $u_j$ and $v_j$ are 
    defined in \cite{Catani:1996vz,Catani:2002hc,Schumann:2007mg}.
    \label{tab:psvardef}
  }
\end{table}

The default parton shower of \Sherpa, dubbed \textsc{CSShower} \cite{Schumann:2007mg},
is based on Catani--Seymour dipole
factorisation~\cite{Catani:1996vz,Catani:2002hc}.  Each branching of an emitter 
parton into two daughters is witnessed by a spectator parton, which takes the
recoil, and ensures that on-shell states are transferred into on-shell states 
and energy-momentum conservation is respected simultaneously.  The emitter
and spectator partons reside either in the initial-state~(I) or final-state~(F), 
such that four dipole types need to be distinguished: II, IF, FI and FF.  In 
this notation, the first letter refers to emitter, and the second to the 
spectator parton. The no-branching probabilities are given by the four 
corresponding Sudakov form factors
\begin{equation}\label{eq:suddecomp}
  \begin{split}
    \Delta_n(\tvar_2,\tvar_1;\cRPS,\cFPS;\alphaS,\fPDFs)
    \,=\;
    \prod\limits_{\text{type}\,\in\,\{\text{FF},\text{FI},\text{IF},\text{II}\}}
    \Delta_n^\text{type}(\tvar_2,\tvar_1;\cRPS,\cFPS;\alphaS,\fPDFs) \,.
  \end{split}
\end{equation}
They share the common form
\begin{equation}\label{eq:generic-single-dipole-sudakov}
  \begin{split}
    &\Delta_n^\text{type}(\tvar_2,\tvar_1;\cRPS,\cFPS;\alphaS,\fPDFs) \\
    &\;=\exp\left(
	  -\sum_{ij}\sum_{k}
	  \int_{\tvar_2}^{\tvar_1} \diff\tvar
	  \int_{z_-}^{z_+} \diff z\;
	  \alphaS(\cRPS\tvar)\,
	  \KPSp_{ij,k}(\tvar,z)\,
	  \frac{\fPDF{c'}{\tfrac{\eta_{c}}{x}}{\cFPS\tvar}}
	       {\fPDF{c}{\eta_{c}}{\cFPS\tvar}}
	\right)\,,
  \end{split}
\end{equation}
wherein the kinematics of the splitting are given by the default choice 
for $\tvar=Q^2\,y\,z(1-z)$ in the massless case while the $\KPSp_{ij,k}(\tvar,z)$ 
denote the coupling and PDF stripped splitting kernels incorporating the remaining pieces of the 
$\KPS_{ij,k}$ and the Jacobian $J$ of the phase-space 
parametrisation. The precise definitions of the variables for each 
dipole type are given in Table~\ref{tab:psvardef}. It directly follows 
that for FF-type dipole splittings the ratio of PDFs is simply unity. 
Eq.\ \eqref{eq:generic-single-dipole-sudakov} further details the 
dependence on the $\alphaS$ and PDF scale factors $\cRPS$ and $\cFPS$. 
These multiplicative factors as well as their variations are assumed to 
be of order one, such that they do not induce spurious large logarithms. 
The generalisation to the massive case is straightforward and only 
involves generalised definitions of $\tvar$, $x$, $y$ and $z$, cf.\ \cite{Schumann:2007mg}.

\subsection{Reweighting trial emissions}

To numerically integrate Sudakov form factors typically 
the Sudakov Veto Algorithm is used  
\cite{Seymour:1994df,Sjostrand:2006za,Hoeche:2009xc,Platzer:2011dq,Hoeche:2011fd,Lonnblad:2012hz}.
Therein the integrands $\KPS$ found in the Sudakov form factors are
replaced with integrable overestimates $\KPSover$.
This is balanced by only accepting a
proposed emission with probability $\Pacc = \KPS / \KPSover$.
A multiplicative factor in $\KPS$ is therefore equivalent to a
multiplicative factor in $\Pacc$~\cite{Hoeche:2009xc}.
This observation is for example used to apply matrix-element
corrections~\cite{Hoeche:2011fd}, where the splitting kernels are replaced
with a real-emission-like kernel $\RME/\BME$.
This is done a-posteriori, i.e.\ the event weight is multiplied
by $(\RME/\BME)/\KPS$, the emission itself is unchanged.
The same method is also used in the {\small\tt Vincia} parton shower to calculate
uncertainty variations for different scales, finite terms of the antenna functions,
ordering parameters and sub-leading colour corrections~\cite{Giele:2011cb}.
Here we employ this technique to account for variations of the strong-coupling parameter
and the PDFs in the shower evolution of LO and NLO QCD matrix elements.

As has been laid out in the previous section,
the emission kernels $\KPS$ depend linearly on $\alphaS$ and 
on a ratio of parton densities $\fPDF{c'}{\eta_c/x}{\cFPS\tvar}/\fPDF{c}{\eta_c}{\cFPS\tvar}$.
A change of PDFs $f \rightarrow \tilde{f}$, the strong coupling
$\alphaS \rightarrow \alphaSt$ and the scale prefactors entering both, i.e.\ 
$\cRPS\to\cRtPS$ and $\cFPS\to\cFtPS$, is equivalent to modifying the 
emission probability accordingly\footnote{ Although the emission scales can not
be reweighted themselves using the presented method, the input scales of
the strong coupling and the PDFs can be changed, as indicated in the text.
We focus on constant prefactors here, but the functional form can also be
changed, although the overall functional form of $\cRPS\tvar$ should be 
restricted to the CMW-like rescaling \cite{Catani:1990rr}.}:
\begin{equation}
  \Pacc \rightarrow \qacc\,\Pacc\,,\qquad
  \qacc \equiv
  \frac{\alphaSt(\cRtPS\tvar)}{\alphaS(\cRPS\tvar)}\;
  \frac{\fPDFt{c'}{\tfrac{\eta_c}{x}}{\cFtPS\tvar}}
       {\fPDF{c'}{\tfrac{\eta_c}{x}}{\cFPS\tvar}}\;
  \frac{\fPDF{c}{\eta_c}{\cFPS\tvar}}
       {\fPDFt{c}{\eta_c}{\cFtPS\tvar}}\,,
  \label{eq:Pacc}
\end{equation}
where the scale dependence and the definition of $\eta_c$ and $x$ can be read 
off the Sudakov form factors given in 
eq.\ \eqref{eq:generic-single-dipole-sudakov} and Table~\ref{tab:psvardef}. 
In case of FF dipoles eq.\ \eqref{eq:Pacc} simplifies significantly as 
the ratios of PDF factors reduces to unity. It further follows,
that the event weight for each accepted
emission needs to be multiplied by the corresponding factor $\qacc$ in order 
to incorporate the new choice of $\alphaS$, PDFs and the scales they are 
evaluated at. Accordingly, the probability to reject an emission is changed to 
\begin{equation}
  \Prej = 1 - \Pacc \rightarrow 1 - \qacc \Pacc
  = \left[ 1 + \left( 1 - \qacc \right) \frac{\Pacc}{1-\Pacc}
    \right]\,\Prej
  \equiv \qrej \,\Prej\,.
  \label{eq:Prej}
\end{equation}
Consequently, for each rejected emission the event weight receives 
a corrective weight of $\qrej$. Proofs that this treatment indeed 
results in the correct Sudakov form factors can be found in 
\cite{Hoeche:2009xc,Mrenna:2016sih,Bellm:2016voq}.

\subsection{Next-to-leading-order matching}\label{sec:nlops}
To match NLO QCD parton-level calculations with subsequent
parton-shower evolution \Sherpa employs a variant of the
original \MCatNLO algorithm presented in \cite{Frixione:2002ik}, 
referred to as \SMCatNLO \cite{Hoeche:2011fd}. Schematically, 
such a \SMCatNLO calculation has the following structure:
\begin{equation}\label{eq:MCatNLO}
  \begin{split}
    \obs^\text{\NLOPS}
    \,=\;&\int\diff\Phi_B\;
          \left[\, \BME(\Phi_B) + \VIME(\Phi_B)
                   +\int\diff x_{a/b}^\prime\,\KPME(\Phi_B,x_{a/b}^\prime)
          \right. \\
         &{} \qquad\qquad
          \left.
          +\sum_j \int \text{d}\Phi_1^j\,
           \left( \DAMEi{j} - \DSMEi{j} \right)(\Phi_B\cdot\Phi_1^j)
          \right.\bigg]\,\PSnlops(O,\Phi_B)\\
         &{}
          +\int\diff\Phi_R\;
           \left[\vphantom{\sum_j}\,
                 \RME(\Phi_R) - \sum_j \DAMEi{j}(\Phi_{B,j}\cdot\Phi_1^j)
           \right]\,
           \PS(O,\Phi_R)\\
    \,=\;&\int\diff\Phi_B\;
          \BbarME(\Phi_B)\;\PSnlops(O,\Phi_B)
          +\int\diff\Phi_R\;
           \HAME(\Phi_R)\;\PS(O,\Phi_R)\;.
  \end{split}
\end{equation}
Here the real-emission contribution $\RME$ of the NLO calculation 
has effectively been split into an infrared-singular (soft) and 
an infrared-regular (hard) part, the resummation kernel $\DAME$ 
and the finite hard remainder $\HAME$, respectively,
such that $\RME=\DAME+\HAME$ ~\cite{Alioli:2008gx,Hoeche:2011fd}. 
The $\BbarME$-function has the following explicit parameter dependences
\begin{equation}\label{eq:Bbar-function}
  \begin{split}
    \BbarME(\Phi_B)
    \,\equiv&\;\BbarME(\Phi_B;\alphaS,\fPDFs;\muR,\muF)\vphantom{\int}\\
    \,=&\;\BME(\Phi_B;\alphaS,\fPDFs;\muR,\muF)
          +\VIME(\Phi_B;\alphaS,\fPDFs;\muR,\muF)\\
       &{}
          +\int\diff x_{a/b}^\prime\,\KPME(\Phi_B,x_{a/b}^\prime;\alphaS,\fPDFs;\muR,\muF)\\
       &{}+\sum_j \int \text{d}\Phi_1^j\;
           \left(
             \DAMEi{j}-\DSMEi{j}
           \right)(\Phi_B\cdot\Phi_1^j;\alphaS,\fPDFs;\muR,\muF)\;.
  \end{split}
\end{equation}
From the perspective of parameter reweighting, the resummation kernel 
$\DAME$ behaves the same way as the subtraction term $\DSME$. In fact, 
in our reweighting implementation the $(\DAME-\DSME)$ contribution is 
treated as a single term, as indicated. It is only to note that their 
PDFs are evaluated at the partonic 
momentum fraction $x_{a/b,j}$ and external flavours $a_j$ and $b_j$ of 
their $\Phi_B\cdot\Phi_1^j$ phase-space configuration rather than 
those of $\Phi_B$. The other parts of the $\BbarME$-function can then be 
reweighted as described in Sec.~\ref{sec:fo_nlo}, leading to
\begin{equation}
  \begin{split}
    \lefteqn{\hspace*{-5mm}\BbarME(\Phi_B;\alphaSt,\fPDFst;\muRt,\muFt)\vphantom{\int}}\\
    \,=\;&\alphaSt^n(\muRtsq)\;\fPDFt{a}{x_a}{\muFtsq}\;\fPDFt{b}{x_b}{\muFtsq}\;\\
         &\times\;
          \left[\vphantom{\int}
            \BMEp(\Phi_B)
            +\alphaSt(\muRtsq)
             \left(\VIMEp(\Phi_B)+\cRp{0}\lR+\tfrac{1}{2}\,\cRp{1}\lR^2\right)
          \right.\\
         &\left.\hspace*{21mm}{}
            +\alphaSt(\muRtsq)
             \int\diff x_{a/b}^\prime\,
             \KPMEp_j(\Phi_j,x_{a/b}^\prime;\fPDFst;\muFcoret)
           \right]\\
         &{}
          +\sum_j\int\diff\Phi_1^j\;
            \fPDFt{a_{j}}{x_{a,j}}{\muFtsq}\;\fPDFt{b_j}{x_{b,j}}{\muFtsq}\;
            \alphaSt^{n+1}(\muRtsq)\;
	  \left[\DAMEpi{j}-\DSMEpi{j}\right](\Phi_B\cdot\Phi_1^j)\;.\hspace*{-15mm}
  \end{split}
\end{equation}
The $\HAME$-function then transforms as 
\begin{equation}\label{eq:def-H}
  \begin{split}
    \lefteqn{\hspace*{-5mm}\HAME(\Phi_R;\alphaSt,\fPDFst;\muRt,\muFt)}\\
    =\;&\RME(\Phi_R;\alphaSt,\fPDFst;\muRt,\muFt)
	 -\sum_j\DAMEi{j}(\Phi_{B,j}\cdot\Phi_{1,j}^j;\alphaSt,\fPDFst;\muRit{j},\muFit{j})\\
    =\;&\alphaSt^{n+1}(\muRtsq)\;\fPDFt{a}{x_a}{\muFtsq}\;\fPDFt{b}{x_b}{\muFtsq}\;
        \RMEp(\Phi_R)\\
       &{}
	 -\sum_j\alphaSt^{n+1}(\muRitsq{j})\;\fPDFt{a}{x_a}{\muFitsq{j}}\;\fPDFt{b}{x_b}{\muFitsq{j}}\;
	  \DAMEpi{j}(\Phi_{B,j}\cdot\Phi_{1,j}^j)\;,
  \end{split}
\end{equation}
wherein each subtraction term $\DAMEi{j}$ has its own scales $\muRi{j}$,
$\muFi{j}$ defined on its underlying Born configuration $\Phi_{B,j}$.
Writing eq.~\eqref{eq:MCatNLO} as a Monte-Carlo sum over events with 
B-like and R-like structure, which are conventionally called $\mathds{S}$ 
and $\mathds{H}$ events in \MCatNLO calculations, and with 
$N=N_\mathds{S}+N_\mathds{H}$, we obtain
\begin{equation}
  \begin{split}
    \obs^\text{\NLOPS}
    \,=&\;\lim\limits_{N\to\infty} \frac{1}{\Ntrial}
          \left\{
                  \sum\limits_{i=1}^{N_\mathds{S}}\;
                  \BbarME(\Phi_{B,i})\;\PSnlops(O,\Phi_{B,i})
          \right.\\
       &\;\left.\hspace*{23mm}{}
                  +\sum\limits_{i=1}^{N_\mathds{H}}\;
                   \HAME(\Phi_{R,i})\;\PS(O,\Phi_{R,i})
          \right\}\,.
  \end{split}
\end{equation}
Thus, under $\muR\to\muRt$, $\muF\to\muFt$, $\alphaS\to\alphaSt$ and 
$\fPDFs\to\fPDFst$ both $\BbarME$ of the $\mathds{S}$-events and~$\HAME$
of the $\mathds{H}$-events transform as composite objects 
in terms of their constituents, as defined above. This leaves the 
\SMCatNLO parton shower, $\PSnlops$, defined through
\begin{equation}\label{eq:PSNLOaction}
  \begin{split}
    \PSnlops(O,\Phi_B) 
    \equiv&\;
          \PSnlops(O,\Phi_B;\cRPS,\cFPS;\alphaS,\fPDFs;\muQsq)\\
    =&\;  \overline{\Delta}_n(\tIR,\tvar^\prime)\,
          O(\Phi_n)\vphantom{\int_{\tvar}^{\tvar}}\\
     &{}
         +\int_{\tIR}^{\tvar^\prime}
	  \diff\Phi_1\;
	  \frac{\DAME(\Phi_B\cdot\Phi_1)}{\BME(\Phi_B)}\;
	  \overline{\Delta}_n(\tvar,\tvar^\prime)\,
	  \PS(O,\Phi_{n+1};\tvar)\,.
  \end{split}
\end{equation}
It differs from the usual $\PS$ of 
eq.\ \eqref{eq:PSaction} with respect to the splitting kernel for the 
first emission and the associated definition of the Sudakov form factor, 
cf.\ \cite{Hoeche:2011fd,Hoche:2012wh}. However, for the 
purpose of reweighting, all trial emissions can be treated in the same way 
as in the standard parton shower as the parameter dependences are identical.

\subsection{Validation}
\label{sec:psvalidation}
To validate the reweighting of scale and parameter dependences 
in \CSShower and \SMCatNLO calculations within the \Sherpa framework we
perform closure tests between reweighting results and dedicated
simulations. 

Our implementation allows to constrain the maximum number of reweighted 
shower emissions per event. For a pure leading-order parton-shower run
or $\mathds{H}$-like events in \SMCatNLO calculations this amounts to 
setting $n_\text{PS}\in\{0,1,2,\ldots,\infty\}$. When considering \SMCatNLO simulations
in addition the parameter $n_\NLOPS\in\{0,1\}$ can be used to disable the 
reweighting of the $\cal{O}(\alphaS)$ emission for $\mathds{S}$-events.  

Of course the reweighting result will only coincide with a dedicated 
calculation if all emissions are reweighted, i.e. $n_\NLOPS=1$ and 
$n_\text{PS}=\infty$. However, by subsequently enabling the reweighting of more 
and more emissions the relevance of their dependences for the determination 
of the full uncertainty can be studied. A finite value of $n_\text{PS}$ can also 
be useful in production, if the effect of reweighting higher-order emissions 
becomes negligible.  The reduced amount of reweighting per event then allows 
for a faster event generation. An additional benefit would be that rare 
high-multiplicity shower histories do not spoil the statistical convergence 
of the reweighted result, even if their exact kinematics might be irrelevant 
for the studied observable.

\subsubsection*{The final-state only case: Thrust in
\HepProcess{\Ppositron\Pelectron\to\Pquark\APquark} events}

To validate \LOPS reweighting, we consider two observables, which are
complementary in their sensitivity to parton-shower emissions. At first, we
consider the event-shape variable thrust $T$ \cite{Farhi:1977sg} in hadronic 
events in $e^+e^-$-collisions at ${\sqrt{s}=\SI{91.2}{GeV}}$.  In this case
QCD emissions are restricted to the final-state. Accordingly, there appear no
PDF factors in the shower reweighting, cf.\ eq.\ \eqref{eq:Pacc}, and thus no
factorisation scale dependence. Moreover, as we consider the leading-order
matrix element for $e^+e^-\to q\bar{q}$ only, the renormalisation scale is also
absent in the hard-process component.  Therefore, we can concentrate on the
pure $\alphaS$ uncertainty in the parton shower here. Leaving the 
perturbative order of its running invariant it is defined by its value at 
the input scale $m_Z$.

\begin{figure}
  \centering
  \includegraphics[]{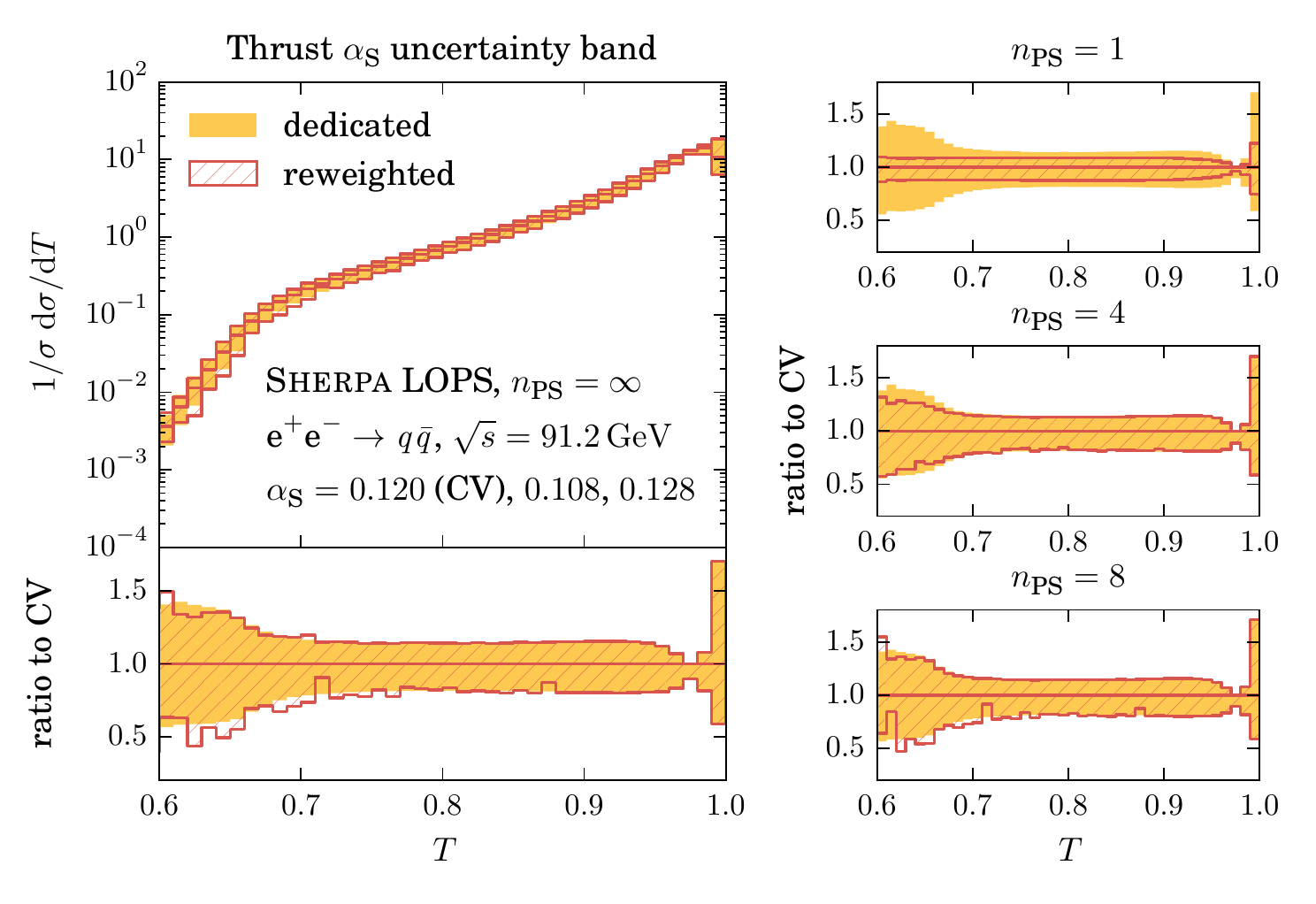}
  \caption{Uncertainty band for the Thrust event shape in dijet production in
      $\Pelectron\Ppositron$ annihilation for a variation of $\alphaS$. The left
      panel shows the nominal distribution and the ratio to the central value.
      The uncertainty band calculated using reweighting (including all
      emissions, i.e. $n_\text{PS}=\infty$) is compared to the one obtained
      from dedicated calculations.  The comparison is repeated in the three
      panels on the right for different maximum number of reweighted emission
      $n_\text{PS}$.}
  \label{fig:lops-thrust}
\end{figure}

In Fig.~\ref{fig:lops-thrust}, we compare $\alphaS$ uncertainty bands generated
by reweighting the nominal prediction with the one generated by dedicated
predictions for each variation.  As in Fig.~\ref{fig:nlo-w-pt}, the uncertainty
band is defined as the envelope over the distributions with different
$\alphaS(m_Z^2)$ input values.  The nominal value is taken as $\alphaS(m_Z^2)=0.120$,
and its up/down variations are $0.128$ and $0.108$, respectively.  Reweighting
bands are presented for $n_\text{PS}=1, 4, 8, \infty$.  The $n_\text{PS}=1$
band underestimates the uncertainty, especially for $T\leq2/3$, where multiple
hard emissions are required, and for $T \approx 1$, the region sensitive to
multiple soft emissions.  For $n_\text{PS}=4$, the uncertainty is underestimated
only for bins with $T\leq2/3$, and less so than for the $n_\text{PS}=1$ case.
The difference between the two choices of $n_\text{PS}=8$ and $\infty$ is
merely statistical and both reproduce the dedicated result very accurately.

\begin{figure}[p]
  \centering
  \includegraphics[]{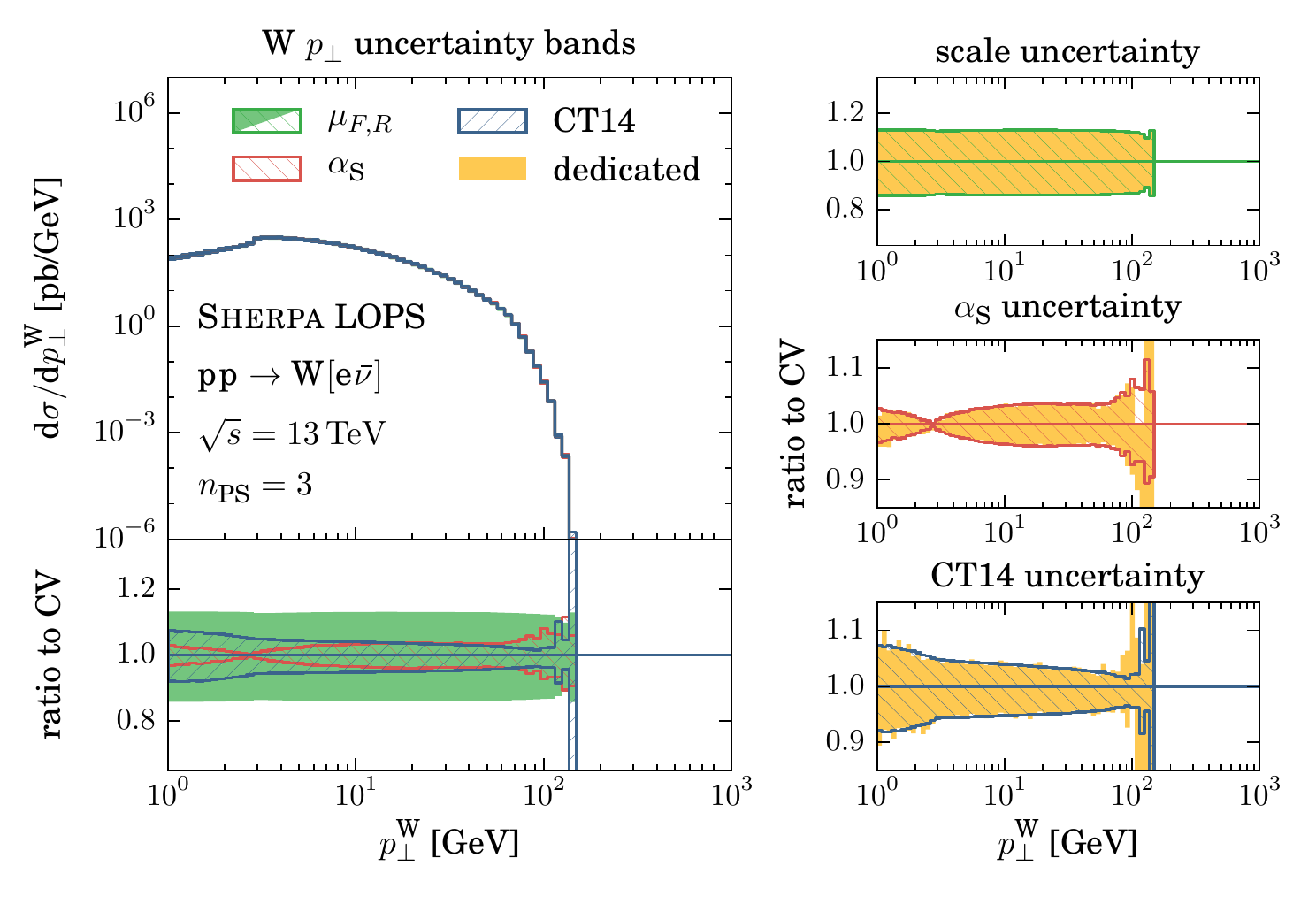}
  \includegraphics[]{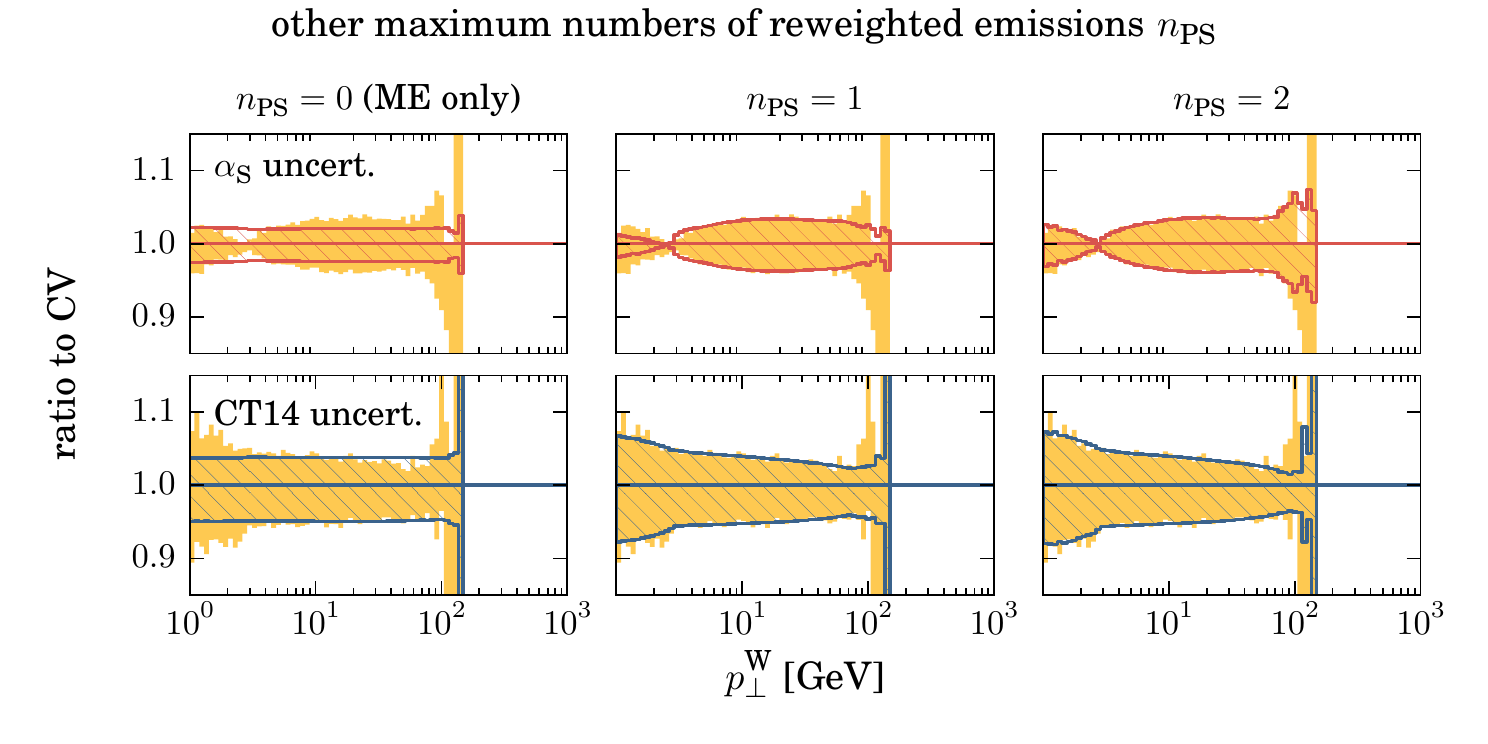}
  \caption{The same as in Fig.~\ref{fig:nlo-w-pt}, but for LO + parton-shower
      (PS) generation.  The uncertainty bands are calculated by reweighting the
      ME and up to $n_\text{PS}$ shower emissions.  In the upper four plots, 
      $n_\text{PS}=3$.  In the lower plots, $n_\text{PS}$ is varied for comparison.  
      The scale uncertainties do not change with $n_\text{PS}$ and are therefore 
      not repeated.}
  \label{fig:lops-w-pt}
\end{figure}

However, for low values of $T$, the statistical fluctuations of the reweighting
results with higher $n_\text{PS}$ grow larger, corresponding to a widening of
the distribution of reweighting factors. Low values of the thrust observable
correspond to the emission of several hard partons, which is less probable in
the parton-shower approximation, and more appropriately modelled in multijet-merged 
calculations, cf.\ Sec.\ \ref{sec:multijet_merging}. In this phase-space
region it is difficult for the reweighting to compensate the multitude of
accepted \emph{soft} emissions off these hard legs, that turn unstable for
$P_\text{acc}\to 1$, with rejected ones, cf.\ eq.~\eqref{eq:Prej}.  This issue
can be addressed by introducing a prefactor for the over-estimator
function~$\KPSover$ in the reweighting runs, to ensure that $P_\text{acc}$ does
not approach $1$, cf.\ \cite{Hoeche:2009xc,Mrenna:2016sih,Bellm:2016voq}.  This
renders the Sudakov Veto Algorithm somewhat less efficient, but is shown to
reduce statistical fluctuations in the reweighting.

\subsubsection*{The initial-state dominated case: $p_\perp^{\PW}$ in
\HepProcess{\Pproton\Pproton\to\PW[\Pe\Pnu]} events}

The second observable considered to validate our \CSShower and \SMCatNLO
reweighting implementation is the \PW-boson transverse-momentum
distribution~$p_\perp^{\PW}$ in \SI{13}{\TeV} proton-proton collisions, that
has already been used in Sec.\ \ref{sec:nlo-validation} in the \NLO
case. The definitions for constructing the uncertainty bands used there
are kept the same, and are stated in Table~\ref{tab:variations}.  We now use
the CT14nlo \PDF set, which uses a Hessian error representation at a
\SI{90}{\percent} confidence level~\cite{Dulat:2015mca}.\footnote{ The reason
    for switching from NNPDF to CT14 PDFs is the strict positivity of the
    latter. The \CSShower rejects emissions when negative PDF values are
    involved, a behaviour which spoils the reweighting in regions where the
    original and the target PDF do not have the same sign. The deviations seem to
    be small in practical applications, but here we chose to establish closure
    in a clean context first.}
Therefore the PDF error band will be larger than before, as it now corresponds
to nearly two standard deviations instead of only one.

Considering a hadronic environment, initial-state emissions are present, which
means that our reweighting factors now include PDF double ratios.  In
Fig.~\ref{fig:lops-w-pt}, we compare \LOPS uncertainty bands for scale,
$\alphaS$ and PDF variations, including comparisons between reweighted and
dedicated predictions for a varying maximum number of reweighted shower
emissions $n_\text{PS}$. Before discussing the bands, we observe that the
tail of the $p_\perp^W$ spectrum is not populated, in particular in comparison to
Fig.~\ref{fig:nlo-w-pt}. This is expected, as the \LO configuration is
restricted to $p_\perp^W=0$, such that all other bins are filled through
recoils against parton-shower emissions only.  However, the phase space of 
parton-shower emissions is restricted to the soft region, and therefore 
the \PW boson can not build up a large recoil.

We now turn to the scale uncertainty band---which is entirely due to
factorisation scale variations, because the \LO matrix element is independent
of $\alphaS$ , and therefore the band underestimates the perturbative
uncertainty.  We also observe that the band is nearly flat. As we vary only the
scales of the matrix-element calculation, the constant spread corresponds to
the factorisation-scale uncertainty of the Born configuration at $p_\perp^W=0$,
merely propagating to higher $p_\perp^W$ bins through the parton shower, which
is unaware of the scale variations. In the matrix-element reweighting, we can
guarantee the same phase-space points as in the dedicated run, such that we see
perfect agreement between dedicated and reweighted predictions. We therefore
omit comparisons for different $n_\text{PS}$ for the scale-uncertainty band.

Looking at the $\alphaS$ uncertainty band, we can see that the envelope
constricts at the position of the peak of the distributions. This reflects that
the variation of $\alphaS$ shifts the position of the peak, such that
variations that are below the nominal distribution on the left side of the
peak, are exceeding the nominal distribution on the right side, and vice versa.
Comparing the reweighted prediction to the dedicated one, we find a flat band
for $n_\text{PS}=0$, corresponding to restricting the reweighting to the
fixed-order matrix element. As the \LO calculation is independent of
$\alphaS$, this only reflects the change of the PDFs, which are fitted to
$\alphaS(m_{\PZ})$.  The reproduction of the shape of the $\alphaS$ uncertainty
improves a lot when reweighting up to one emission ($n_{\text{PS}}=1$), and
slightly more when adding another emission on top ($n_{\text{PS}}=2$).

For the PDF uncertainty, we see that the reweighting with $n_\text{PS}=0$
underestimates it by at least 1--\SI{5}{\percent} for small transverse momenta,
and overestimate it around the \PW mass. As for the $\alphaS$ uncertainty, this
improves for $n_\text{PS}=1,2$.

The last depicted step, i.e.\ $n_{\text{PS}}=3$, on the other hand, does not
contribute further to the reproduction of the $\alphaS$ and PDF uncertainties.
No significant differences with respect to the $n_\text{PS}=2$ case is
observed. It can be concluded that it is sufficient to include up to two
emissions to reproduce the uncertainty bands for this observable.

This is to be expected, as the gauge boson recoils against the shower emissions 
and is therefore mostly affected by the few hardest branchings. These mainly 
originate from the incoming hard virtual partons, so the generally softer 
final-state emissions barely contribute.  Although we do not reproduce this 
here, we confirmed this by entirely disabling final-state emissions, which 
showed no effect on the results. 

\begin{figure}
  \centering
  \includegraphics[]{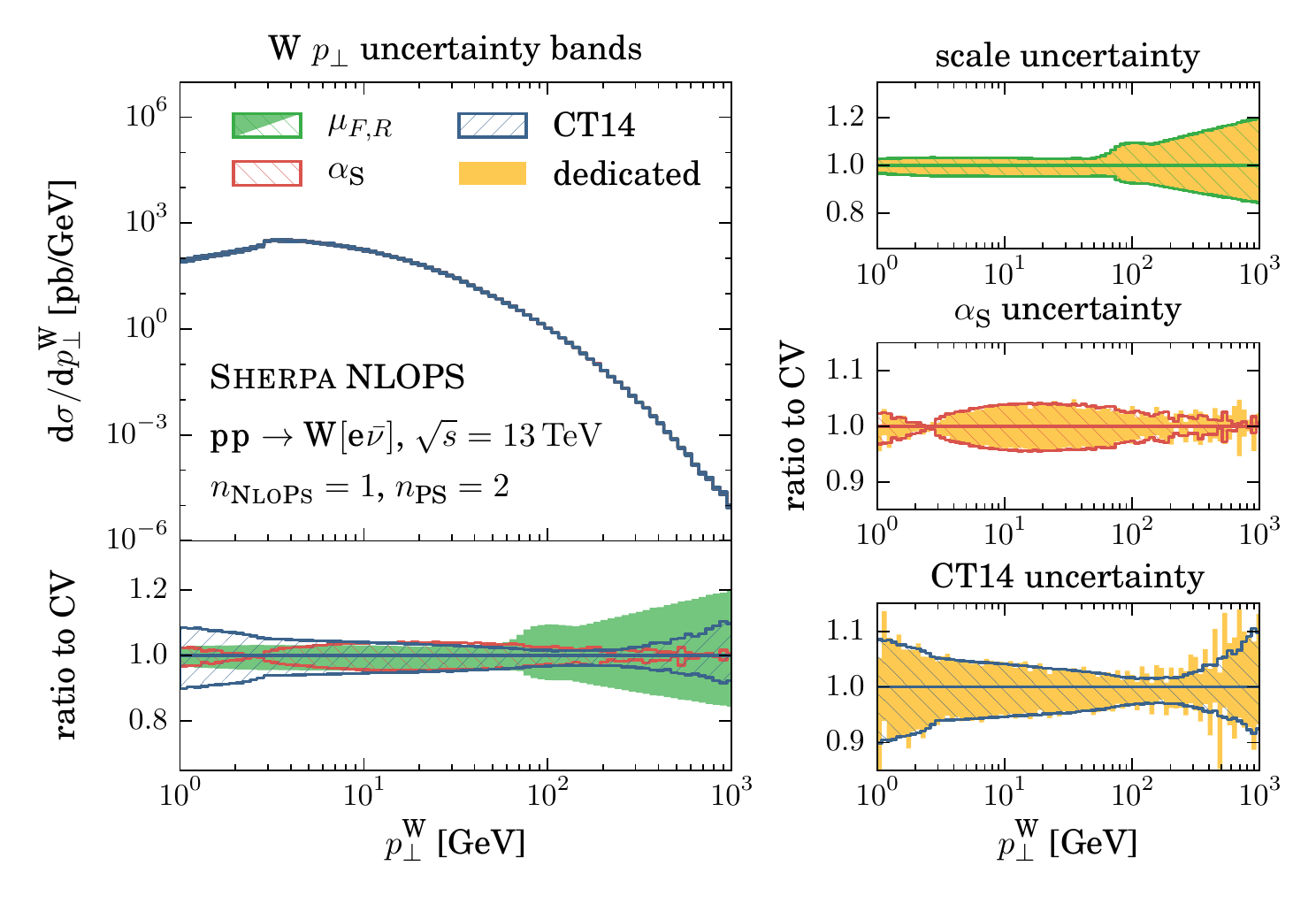}
  \includegraphics[]{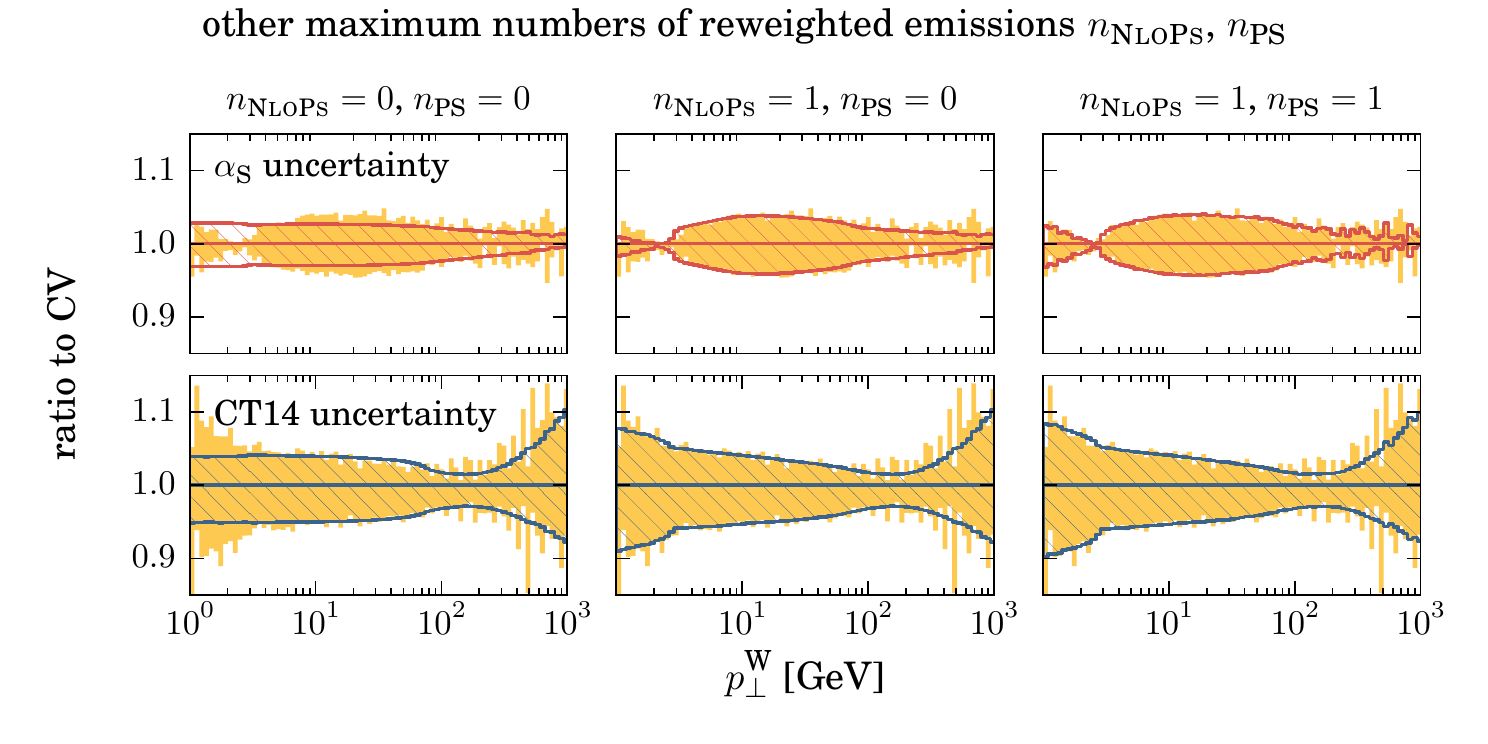}
  \caption{The same as in Fig.~\ref{fig:lops-w-pt}, but for NLO + parton-shower
      (PS) generation.  The uncertainty bands are calculated by reweighting the
      ME and a maximum number of emissions from the \MCatNLO ($n_\NLOPS$) and
      the ordinary PS ($n_\text{PS}$).  $n_\NLOPS$ is constrained to 0 or 1,
      as the \MCatNLO prescription only affects the first emission.  In the 
      upper four plots we consider $n_\NLOPS=1$ and $n_\text{PS}=2$, thus up 
      to three emissions get reweighted.  In the lower plots, we consider
      variations of $n_\NLOPS$ and $n_\text{PS}$.}
  \label{fig:nlops-w-pt}
\end{figure}

In Fig.~\ref{fig:nlops-w-pt} we present the validation of \NLOPS predictions
for the \PW-boson transverse-momentum distribution.  Overall, we get a similar
picture as in the \LOPS case. The main differences are the increased
high-$p_\perp^{\PW}$ reach and the significantly smaller scale uncertainties in
the low $p_\perp^{\PW}$ range, a consequence of including the complete set of
$\cal{O}(\alphaS)$ corrections to the production process. For large
$p_\perp^{\PW}$, the uncertainty increases again, because we fall back to a \LO
description again: \emph{Only} the $2\to 3$ matrix element still contributes.

To assess the quality of the reweighting, we consider again different settings 
for the parameters $n_\NLOPS$ and $n_\text{PS}$. Assuming  $n_\NLOPS=n_\text{PS}=0$,
only the scale variations of the hard process are considered and
the parton-shower contribution to the $\cal{O}(\alphaS)$ correction is {\em not}
reweighted. Furthermore, we present results for $n_\NLOPS=1$ and $n_\text{PS}=0,1,2$.
With these settings, the $\cal{O}(\alphaS)$ corrections get properly reweighted, but the
number of subsequent shower emissions off the $\mathds{S}$- and $\mathds{H}$-like 
events treated correctly is varied. We observe a saturation for reproducing the 
dedicated calculations at $n_\NLOPS + n_\text{PS} \geq 2$, with no further improvement 
when $n_\text{PS}$ is increased from $1$ to $2$. This confirms the findings made when
considering the \LOPS setup in Fig.~\ref{fig:lops-w-pt}: The gauge-boson 
transverse-momentum distribution is dominated by the few hardest emissions.


\section{Reweighting multijet-merged calculations}
\label{sec:multijet_merging}

In this section we address the reweighting of multijet-merged
event generation runs. These approaches allow to combine LO or NLO QCD 
matrix elements of different multiplicity dressed with parton showers 
into inclusive samples. Accordingly, the production of jets associating
a given core process can be modelled through exact matrix elements rather 
than relying on the logarithmic approximation of the parton shower 
only. In particular, when considering hard jet kinematics or angular 
correlations such techniques prove to be indispensable to properly 
describe experimental observations, see for instance 
\cite{Aad:2013ysa,Chatrchyan:2014fsa,Khachatryan:2014uva,Aaboud:2016yus}.

To first approximation the reweighting as described in the
previous sections can be used without change, only that the perturbative 
order~$p$ is no longer a constant across the sample, but varies for each 
event, corresponding to the considered matrix-element parton multiplicity. 
However, there are also new algorithm-specific intricacies which complicate the 
dependence on the input parameters and need to be dealt with to allow for
a consistent reweighting. The LO and NLO merging techniques employed within
the \Sherpa framework are presented in \cite{Hoeche:2009rj} 
and \cite{Hoeche:2012yf,Gehrmann:2012yg}, respectively. They rely on the 
reconstruction of parton-shower histories for multi-parton amplitudes 
that set the parton-shower initial conditions for their subsequent evolution. 
This is achieved by running a {\em backward-clustering} algorithm that 
identifies a corresponding core process and calculates hard-parton splitting 
scales that serve as predetermined shower branchings. In the \Sherpa approach
the actual parton shower then starts off the reconstructed core process
and implements the predetermined hard splittings based on a {\em truncated
shower}. Furthermore, it is the purpose of the truncated-shower evolution
to implement possible Sudakov vetoes for shower emissions above the 
phase-space separation or merging scale $\Qcut$. It should be emphasised 
here, parton-shower reweighting is vital when using modified input 
parameters in order to cancel the $\Qcut$ dependence to the accuracy of 
the parton shower. In case only the hard-process matrix element parameters 
get reweighted, the dependence on $\Qcut$ is cancelled to leading-logarithmic 
accuracy only, however, residual subleading contributions from the running coupling 
or the PDF evolution remain \cite{Hoeche:2012yf}. 

For the reweighting of the truncated-shower Sudakov veto probability the 
methods described in Sec.~\ref{sec:parton_shower} can be applied. In what 
follows we will detail the specifics of the reweighting procedure for LO 
and NLO multijet-merging runs with \Sherpa supplemented by an extensive 
validation of the implementation. 

\subsection{Preliminaries}

\begin{figure}
  \centering
  \begin{picture}(400,100)
    \put(0,0){\includegraphics[width=100pt]{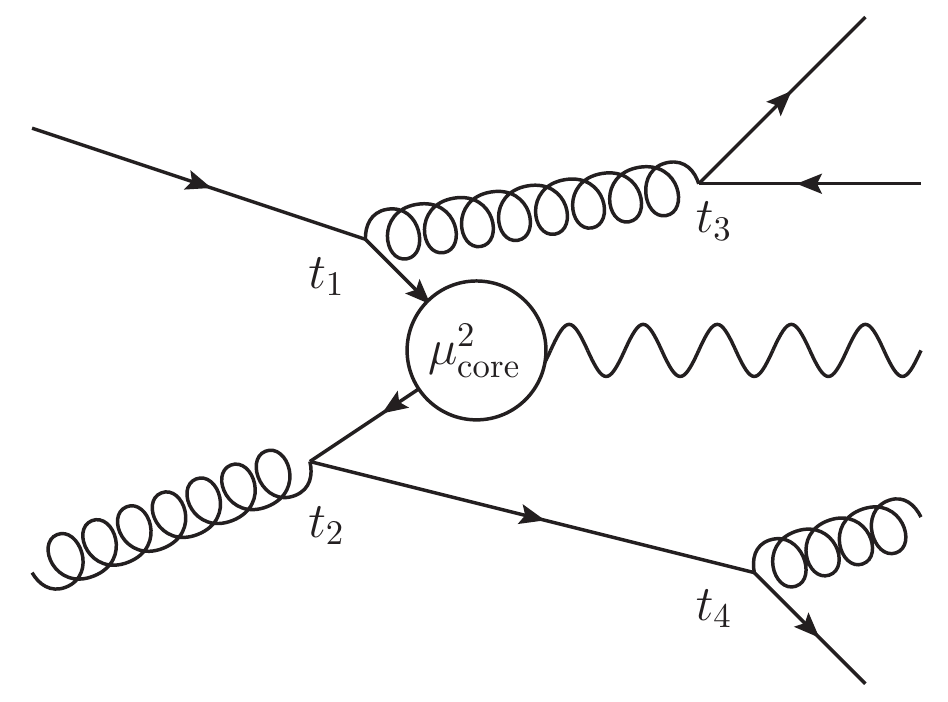}}
    \put(106,48){\scriptsize QCD only}
    \put(115,34){$\Longleftarrow$}
    \put(150,0){\includegraphics[width=100pt]{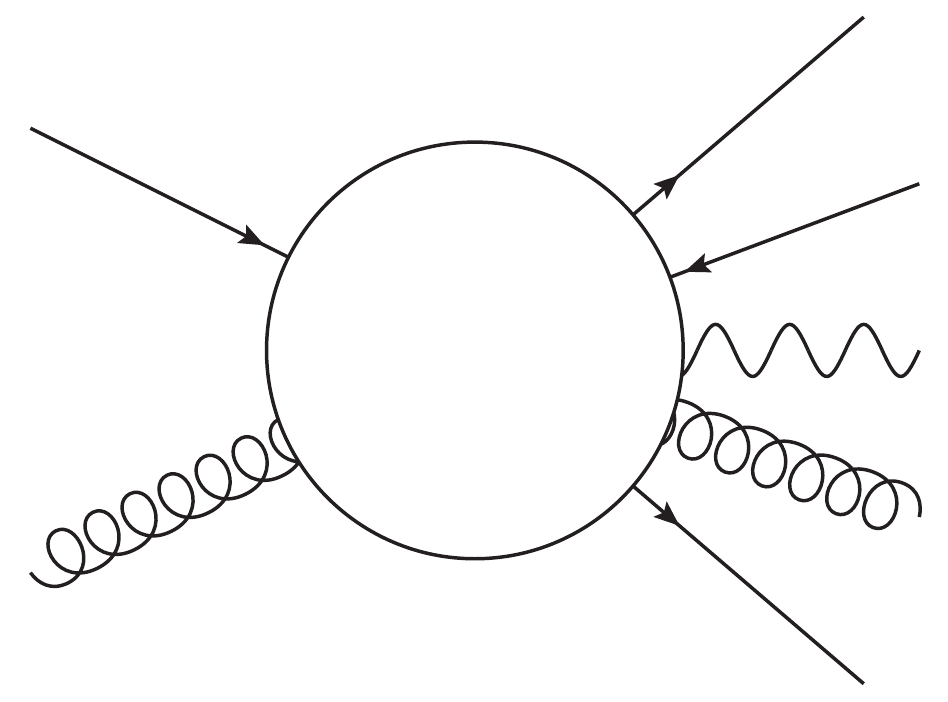}}
    \put(252,48){\scriptsize QCD \& EW}
    \put(265,34){$\Longrightarrow$}
    \put(300,0){\includegraphics[width=100pt]{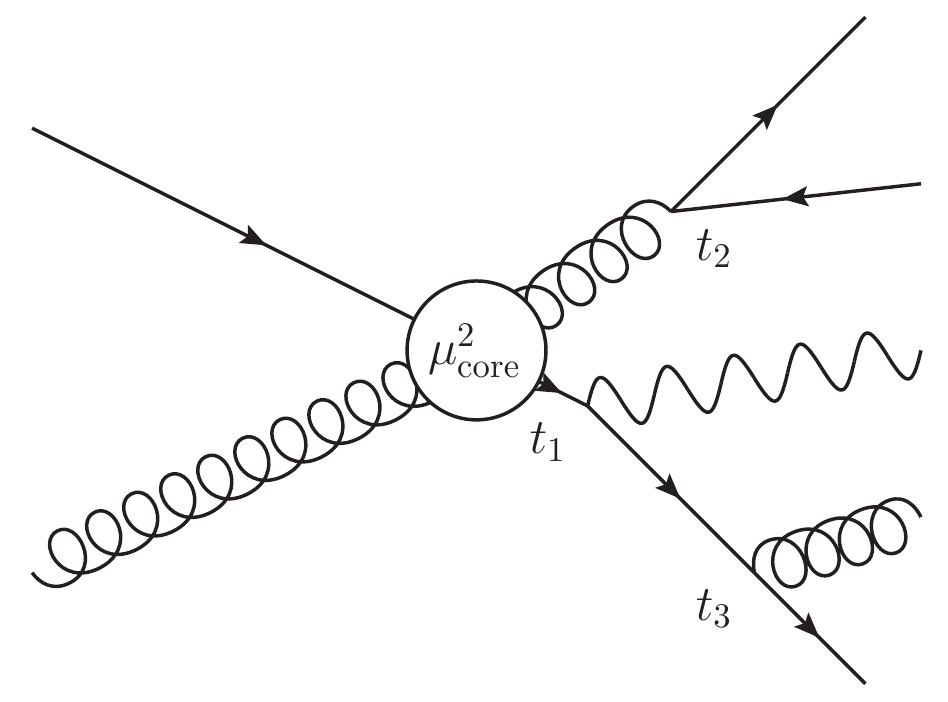}}
  \end{picture}
  \caption{Possible parton-shower histories of a $qg\to Zgqq^\prime\bar q^\prime$ 
           matrix element allowing only QCD splittings (left) and also 
           including electroweak splittings (right).}
  \label{fig:clusterhistos}
\end{figure}
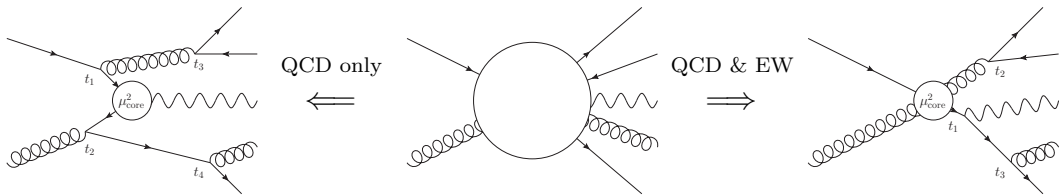

Common to the LO and NLO merging techniques used in \Sherpa, cf. 
\cite{Hoeche:2009rj,Hoche:2010kg,Hoeche:2012yf,Gehrmann:2012yg,Schalicke:2005nv}, is the 
separation of the emission phase space into a soft and a hard 
region, defined through a suitable $m$-parton measure $\Qvar_m$ and a 
separation criterion $\Qcut$. For each parton configuration $\Phi_m$ 
with $\Qvar_m>\Qcut$ a shower history that represents the event as a 
core process with subsequent $1\to 2$ shower splittings is 
probabilistically build through backward clustering. The resulting sequence of 
cluster steps is characterised by tuples $\{a_i,b_i,x_{a,i},x_{b,i},\tvar_i\}$, 
recording the (possibly changing) initial-state flavours and momentum 
fractions as well as the evolution variable of each splitting. 
We allow for both QCD and EW splitting functions \cite{Hoeche:2009rj,Krauss:2014yaa} 
to identify such splitting processes and veto recombinations that would lead 
to the reduction of configurations which are not present in the matrix elements%
\footnote{
  An example here is the interpretation of a \HepProcess{\Ppositron\Pelectron\to
  \Pgluon\Pdown\APdown} configuration. Its matrix element does not 
  contain terms/diagrams that allow the quark-antiquark pair to be 
  clustered.
}. 
Figure \ref{fig:clusterhistos} details possible cluster histories for a given
\HepProcess{\Pp\Pp \to \PZ + 4\,\text{jets}} configuration, depending on its
kinematics, allowing for QCD splittings only (left) or both QCD and EW
splittings (right).

The sequence $\{\tvar_i\}$ of reconstructed branching scales then may 
be either ordered or unordered, with an ordered 
history satisfying $\tvar_j<\tvar_{j-1}<\ldots<\tvar_1<t_0=\muFcoresq$. 
The recombination probabilities in 
each clustering step are determined by the forward-splitting probabilities 
and are therefore dependent on the parton shower and its parameters and 
choices. This is reflected, step-by-step, in the addition of one factor 
of $\alphaS$ (when appropriate) at the reconstructed splitting scale, a 
ratio of PDFs at the reconstructed initial flavours and their momentum 
fractions, and a Sudakov form factor describing the evolution 
of each step. 

In the \Sherpa implementations the $\alphaS$ and PDF factors are added 
explicitly onto the respective matrix elements and can therefore be reweighted 
directly. The Sudakov form factor, on the other hand, is implemented through 
a vetoed truncated parton shower \cite{Hoeche:2009rj,Hoche:2010kg}.  The 
truncated shower itself, accounting for the possibility 
of soft parton-shower emissions between subsequent reconstructed hard 
emissions, i.e.\ with $\tvar_m<\tvar<\tvar_{m-1}$ but $\Qvar<\Qcut$, can 
be reweighted with the methods described in Sec.~\ref{sec:parton_shower}. If, 
however, an emission with $\Qvar>\Qcut$ occurs the event is vetoed. Practically, 
this is accounted for through increasing $\ntrial$ of the next accepted event 
by $\ntrial$ of the vetoed event. Thus, $\ntrial$ becomes dependent on the 
parton-shower parameters.

A special remark concerning unordered histories, i.e.\ histories whose 
sequence of $\{\tvar_i\}$ has at least one pair $\tvar_k\ge\tvar_{k-1}$, 
is in order. Such histories can be encountered in various 
configurations, e.g.\ when the last clustering step produces a splitting 
scale larger than the nominal starting scale of the core process%
\footnote{
  An example here is the interpretation of a \HepProcess{\Pgluon\Pquark\to\PZ\Pquark}
  configuration. In regions of large transverse momenta of the final 
  state parton its identified branching scale $\tvar_1$ is larger than 
  the starting scale $\tvar_0$ of the core process 
  \HepProcess{\Pquark\APquark\to\PZ}, usually defined as the \PZ virtuality.
}
or the flavour structure only allows further clusterings at scales 
$\tvar_{k-1}$ lower than the last identified one $\tvar_k$%
\footnote{
  An example here is the interpretation of a \HepProcess{\Ppositron\Pelectron\to
  \Pdown\APdown\Pup\APup} configuration. In a first step 
  there are only two choices to cluster, resulting in an identified 
  branching scale $\tvar_2$. There now is a finite region in phase 
  space where the gluon can (only) be clustered with scale 
  $\tvar_1<\tvar_2$. 
}.
As such configurations cannot be generated by a strictly ordered parton 
shower, for each unordered step neither the accompanying PDF ratio nor 
Sudakov factor is therefore present in the 
calculation. More than one unordering in a cluster history of a given 
event is possible and in fact likely at high multiplicities. PDF ratios 
and Sudakov factors then of course only occur in the ordered subhistories 
in between the 
unorderings. For the sake of clarity and brevity we will omit this case from 
the discussion of the following subsections. Its implications to the 
algorithm, and therefore to the reweighting, are straightforward. If
ordered histories are enforced, core configurations beyond the standard
$2\to 2$ processes occur. Independent of the presence and number of 
unorderings, the renormalisation and factorisation scales $\mu_R$ and $\mu_F$ 
are always set in the way, as will be detailed below.

\subsection{The leading-order case}
\label{sec:merging-lo}

We start the discussion with the simplest case, where all matrix elements 
used in the merging are given at leading order. A LO multijet-merged 
calculation, with Born matrix elements at $\order{\alphaS^{n+j}}$, containing 
$j$ additional partons relative to the core process, has the following 
structure
\begin{equation}\label{eq:MEPSatLO}
  \begin{split}
    \obs^\text{\MEPSatLO}
    \,=&\;\sum_{j=0}^{j_\text{max}}
          \int\diff\Phi_j\;\BMEmerge_j(\Phi_j)\;
          \Theta(\Qvar_j-\Qcut)\;
          \PStv(O,\Phi_j) \\
    \,=&\;\lim\limits_{N\to\infty}\,\frac{1}{\Ntrial}
          \sum_{i=1}^N\sum_{j=0}^{j_\text{max}}
          \BMEmerge_j(\Phi_{j,i})\;
          \Theta(\Qvar_j-\Qcut)\;
          \PStv(O,\Phi_{j,i})\;.
  \end{split}
\end{equation}
Note that $\Phi_{j}$ here denotes the entire final-state phase space of the
process, including all particles of the core process. As before, $\Qvar_j$ 
is a suitable infrared-safe distance measure of $\Phi_j$. The 
$\Theta$-function thus realises a minimum separation of $\Qcut$ and acts as 
an infrared regulator. $\PStv$ is the vetoed 
truncated parton shower derived from eq.\ \eqref{eq:PSaction}. As the
limit in the second line is well defined, it can be transposed with the
summation over parton multiplicities. As the ingredient leading-order matrix
elements need to incorporate the soft-collinear resummation properties of the
parton shower, they have the following parameter dependences:
\begin{equation}
  \begin{split}
    \BMEmerge_j(\Phi_j)
    \,\equiv&\;\BMEmerge_j(\Phi_j;\alphaS,\fPDFs;\muRcore,\muFcore,\cRPS,\cFPS;
                     \{a_i,b_i, x_{a,i},x_{b,i}, \tvar_i\})\,.
  \end{split}
\end{equation}
The cluster steps $\{a_i, b_i, x_{a,i}, x_{b,i}, \tvar_i\}$ denote the 
identified cluster history of the configuration $\Phi_j$, as discussed 
above. Therein, the $a_i,\,b_i$ are the possibly changing initial-state 
flavours, the $x_{a,i},\,x_{b,i}$ their momentum fractions, and the 
$\tvar_i$ are the reconstructed values of the parton-shower evolution 
variable at each splitting. Together with the $\alphaS$ and PDF scale
prefactors $\cRPS$, $\cFPS$ of the parton shower, the cluster steps
relate $\BMEmerge_j$ to the scale and PDF stripped Born matrix element
$\BMEp_j$ encountered in Sec.~\ref{sec:fo:lo},
\begin{equation}\label{eq:pdfratios-ordered}
  \begin{split}
    \lefteqn{\hspace*{-5mm}\BMEmerge_j(\Phi_j;\alphaS,\fPDFs;\muRcore,\muFcore,
                                       \cRPS,\cFPS;\{a_i,b_i,x_{a,i},x_{b,i},\tvar_i\})}\\
    \,=\;&\prod\limits_{i=1}^j\frac{\fPDF{a_i}{x_{a,i}}{\cFPS\tvar_i}}
                                   {\fPDF{a_{i-1}}{x_{a,i-1}}{\cFPS\tvar_i}}\;
          \fPDF{a_0}{x_{a,0}}{\muFcoresq}
	  \prod\limits_{i=1}^j\frac{\fPDF{b_i}{x_{b,i}}{\cFPS\tvar_i}}
	                           {\fPDF{b_{i-1}}{x_{b,i-1}}{\cFPS\tvar_i}}\;
          \fPDF{b_0}{x_{b,0}}{\muFcoresq}\\
         &\times\;
	  \alphaS^{n+j}(\muRsq)\;\BMEp_j(\Phi_j)\,.
  \end{split}
\end{equation}
In this notation, the core scale is $\tvar_0 = \muFcore^2$, it is therefore not
multiplied by the prefactors of the parton shower. The partonic momentum
fractions of the core process are $x_{a,0},\,x_{b,0}$. 

The scales of each single $\alphaS$ within the cluster history vary, but an
effective global renormalisation scale can be defined through
\begin{equation}\label{eq:meps-as}
  \begin{split}
    \alphaS^{n+j}(\muRsq) = \alphaS^{n+e}(\muRcoresq)\;
                            \prod_{i=1}^{j}\alphaS^{1-\epsilon_i}(\cRPS\tvar_i)\,,
  \end{split}
\end{equation}
where $\epsilon_i=0$ if the identified splitting process at branching $i$ 
is of QCD-type, and $1$ otherwise, $e=\sum_{i=1}^j\epsilon_i$. To consistently 
vary the $\muR$ scale, we consider variations of the splitting scales $\tvar_i$ 
and the core scale $\muRcore$ on the right-hand side by a common factor, solving 
for the prefactor of the effective $\muR$ to be used in the matrix-element 
calculation. Thus, while up to NLO accuracy $\muR$ is varied by the same 
common factor, the full solution of this procedure results in slightly 
larger variations of the effective renormalisation scale.

Apart from the Sudakov form factors the soft-collinear structure of the $\BMEmerge_j$ 
is now identical to the emission of $j$ partons off a $\BME_0$ configuration 
with the parton shower described in section~\ref{sec:parton_shower}. In 
case of final-state splittings the ratio of parton distribution functions 
is simply unity as neither the partonic $x_{a/b,i}$ and $x_{a/b,i-1}$ 
nor the initial-state flavours $a_i,\,b_i$ and $a_{i-1},\,b_{i-1}$ differ.
In principle, with every ratio of PDFs there 
is also a ratio of flux factors. However, all such factors cancel except for 
the outermost ones, corresponding to $\Phi_j$ and, hence, are regarded as 
part of $\BMEp_j$.

Changing the scales $\muRcore\to\muRcoret$, $\muFcore\to\muFcoret$, 
$\cRPS\to\cRtPS$, $\cFPS\to\cFtPS$ 
as well as $\alphaS\to\alphaSt$ and $\fPDFs\to\fPDFst$ results in
\begin{equation}\label{eq:MEPS-scale-var-ordered}
\begin{split}
\lefteqn{\hspace*{-5mm}\BMEmerge_j(\Phi_j;\alphaSt,\fPDFst;\muRcoret,\muFcoret,
                                    \cRtPS,\cFtPS;\{a_i,b_i,x_{a,i},x_{b,i},\tvar_i\})}\\
\,=\;&\prod\limits_{i=1}^j\frac{\fPDFt{a_i}{x_{a,i}}{\cFtPS\tvar_i}}
                                {\fPDFt{a_{i-1}}{x_{a,i-1}}{\cFtPS\tvar_i}}\;
        \fPDFt{a_0}{x_{a,0}}{\muFcoretsq}
        \prod\limits_{i=1}^j\frac{\fPDFt{b_i}{x_{b,i}}{\cFtPS\tvar_i}}
                                {\fPDFt{b_{i-1}}{x_{b,i-1}}{\cFtPS\tvar_i}}\;
        \fPDFt{b_0}{x_{b,0}}{\muFcoretsq}\\
        &\times\;
        \alphaSt^{n+j}(\muRtsq)\;\BMEp_j(\Phi_j)\,.
\end{split}
\end{equation}
The scale $\muRtsq$ is now calculated from eq.\ \eqref{eq:meps-as} using 
$\muRcoretsq$ and $\cRtPS$ as input.

Eq.\ \eqref{eq:MEPS-scale-var-ordered} describes what happens 
to the matrix-element part of a multijet-merged calculation. This 
leaves the vetoed truncated shower $\PStv$. While the truncated and standard 
shower part is described in section~\ref{sec:parton_shower}, 
the vetoed shower leads to vetoed events. As vetoed events correspond 
to events whose weights have been set to zero, their description is 
equivalent to increasing the number of trials, $\ntrial$, by one. 
Thus, when varying the parameters of the parton shower, also the 
probabilities of vetoing events are changed. Consequently, 
$\ntrial$ acquires a dependence on the parameters of the variation. 
Thus, now explicitly stating the dependence on the shower's starting 
scale $\muQsq$ and the merging scale $\Qcut$, 
\begin{equation}
  \begin{split}
    \lefteqn{\hspace*{-5mm}\ntrial(\muQsq;\Qcutsq;\{a_i,b_i,x_{a,i},x_{b,i},\tvar_i\})}\\
    \,\equiv&\;\ntrial(\muQsq;\Qcutsq;\{a_i,b_i,x_{a,i},x_{b,i},\tvar_i\};
                       \alphaS,f;\cRPS,\cFPS)\\
    \,=&\;\frac{1-\Delta_{n+j}(\tIR,\muQsq;\Qcutsq;\alphaSt,\fPDFst;
                               \cRtPS,\cFtPS;\{a_i,b_i,x_{a,i},x_{b,i},\tvar_i\})}
               {1-\Delta_{n+j}(\tIR,\muQsq;\Qcutsq;\alphaS,\fPDFs;
                               \cRPS,\cFPS;\{a_i,b_i,x_{a,i},x_{b,i},\tvar_i\})}\,,
  \end{split}
\end{equation}
where
\begin{equation}
  \begin{split}
    \lefteqn{\hspace*{-5mm}\Delta_{n+j}(\tIR,\muQsq;\Qcutsq;\alphaS,\fPDFs;
                                        \cRPS,\cFPS;\{a_i,b_i,x_{a,i},x_{b,i},\tvar_i\})}\\
    \,=&\;\prod\limits_{i=1}^j
          \exp\left\{
               -\int\limits_{\tvar_i}^{\tvar_{i-1}}\diff\Phi_1\;
	            \KPS_{n+i}(\Phi_1;\cRPS,\cFPS;\alphaS,\fPDFs)\;
	            \Theta(\Qvar_{n+i}>\Qcut)
              \right\}
    \;.
  \end{split}
\end{equation}
Thus, $\ntrial$ corresponds to the survival probability between the 
unfolding of preexisting splittings when evolving from the $n$-parton core 
configuration to the $(n+j)$-parton configuration. Hence, when changing the 
parameters of the simulation the truncated showers probability to emit a 
parton with $\Qvar>\Qcut$ must be re-evaluated following the substitutions 
$\cRPS\to\cRtPS$, $\cFPS\to\cFtPS$, $\alphaS\to\alphaSt$ and 
$\fPDFs\to\fPDFst$ using the methods of Sec.\ \ref{sec:parton_shower}. 
Note, all emissions produced by the truncated shower {\em prior} to the 
one that triggers the veto need to be reweighted as they impact the 
initial conditions for that emission.

\subsection{The next-to-leading-order case}

The merging of multijet matrix elements at next-to-leading order accuracy 
proceeds schematically similar as in the leading order case. The input 
quantity is now the \NLOPS matched $(n+j)$-parton configuration, thus
\begin{equation}\label{eq:MEPSatNLO}
  \begin{split}
    \obs^\text{\MEPSatNLO}
    \,=&\;\sum_{j=0}^{j_\text{max}}
          \left[
                \int\diff\Phi_j\;
                \BbarMEmerge_j(\Phi_{j})\;
                \Theta(\Qvar_j-\Qcut)\;
                \PStvnlops(O,\Phi_j)
          \right.\\
       &  \left.\hspace*{10mm}{}
                +\int\diff\Phi_{j+1}\;
                 \HAMEimerge{j}(\Phi_{j+1},\Qcut)\;
                 \PStv(O,\Phi_{j+1})
          \right] \\
    \,=&\;\lim\limits_{N\to\infty}\,\frac{1}{\Ntrial}
          \left\{
                  \sum\limits_{i=1}^{N_\mathds{S}}\sum\limits_{j=0}^{j_\text{max}}\;
                  \BbarMEmerge_j(\Phi_{j,i})\;
                  \Theta(\Qvar_j-\Qcut)\;
                  \PStvnlops(O,\Phi_{j,i})
          \right.\\
        & \left.\hspace*{23mm}{}
                 +\sum\limits_{i=1}^{N_\mathds{H}}\sum\limits_{j=0}^{j_\text{max}}\;
                  \HAMEimerge{j}(\Phi_{j+1,i},\Qcut)\;
                  \PStv(O,\Phi_{j+1,i})
          \right\}\,,\hspace*{-20mm}
  \end{split}
\end{equation}
keeping the notation of eqs.\ \eqref{eq:MCatNLO} and \eqref{eq:MEPSatLO}. 
For the $\mathds{S}$-events the same $\Theta$-function of eq.\ 
\eqref{eq:MEPSatLO} is used as an infrared regulator and the \SMCatNLO 
parton shower of eq.\ \eqref{eq:PSNLOaction} is replaced by its 
vetoed version, it only matches the softest emission in $\tvar$ and, thus, 
does not generate truncated emissions. These are added dressing it with 
additional emissions through the standard parton shower. 
As all ingredients of $\BbarMEmerge_j$ are evaluated at the same phase-space 
point $\Phi_j$ they share a common cluster history 
$\{a_i,b_i,x_{a,i},x_{b,i},$ $\tvar_i\}$. Hence, again suppressing any 
further $\Qcut$-dependence which is not varied,
\begin{equation}
  \begin{split}
    \BbarMEmerge_j(\Phi_j)
    \equiv&\;\BbarMEmerge_j(\Phi_j;\alphaS,\fPDFs;\muRcore,\muFcore,\cRPS,\cFPS;
                            \{a_i,b_i, x_{a,i},x_{b,i},\tvar_i\})
  \end{split}
\end{equation}
transforms under the replacements $\cRPS\to\cRtPS$, $\cFPS\to\cFtPS$, 
$\alphaS\to\alphaSt$ and $\fPDFs\to\fPDFst$ in the following way:
\begin{equation}
  \begin{split}
    \lefteqn{\hspace*{-5mm}\BbarMEmerge_j(\Phi_j;\alphaSt,\fPDFst;
                                          \muRcoret,\muFcoret,\cRtPS,\cFtPS;
                                          \{a_i,b_i, x_{a,i},x_{b,i},\tvar_i\})}\\
    =\;&\prod\limits_{i=1}^j\frac{\fPDFt{a_i}{x_{a,i}}{\cFtPS\tvar_i}}
                                 {\fPDFt{a_{i-1}}{x_{a,i-1}}{\cFtPS\tvar_i}}\;
        \fPDFt{a_0}{x_{a,0}}{\muFcoretsq}\\
       &\quad\times
	\prod\limits_{i=1}^j\frac{\fPDFt{b_i}{x_{b,i}}{\cFtPS\tvar_i}}
	                         {\fPDFt{b_{i-1}}{x_{b,i-1}}{\cFtPS\tvar_i}}\;
        \fPDF{b_0}{x_{b,0}}{\muFcoretsq}\\
       &\quad\times\;
	\alphaSt^{n+j}(\muRtsq)\;
	\left[
          \BMEp_j(\Phi_j)
          +\alphaSt(\muRtsq)
           \left(
             \VIMEp_j(\Phi_j)+\cRpi{j}{0}\lR+\tfrac{1}{2}\,\cRpi{j}{1}\lR^2
           \right)\vphantom{\int}
        \right.\\
       &\left.\hspace*{42.7mm}{}
          +\alphaSt(\muRtsq)
           \int\diff x_{a/b}^\prime\,
           \KPMEp_j(\Phi_j,x_{a/b}^\prime;\fPDFst;\muFcoret)
         \right.\\
       &\left.\hspace*{42.7mm}{}
          +\alphaSt(\muRtsq)\sum_k\int\text{d}\Phi_1^k
           \left(
             \DAMEpi{k}
             -\DSMEpi{k}
           \right)(\Phi_j\cdot\Phi_1^k)
        \right.\bigg]\hspace*{-10mm}\\
       &{}
        -\sum_{i=1}^j\frac{\alphaSt(\muRtsq)}{2\pi}\,
         \log\frac{\tvar_{i-1}}{\tvar_i}
         \left(
	   \sum\limits_{c=q,g}\int\frac{\diff x_{a,i}^\prime}{x_{a,i}^\prime}\;
	   P_{ac}(x_{a,i}^\prime)\,\fPDFt{c}{\tfrac{x_{a,i}}{x_{a,i}^\prime}}{\cFtPS\tvar_i}
	 \right.\\
       &\left.\hspace*{42mm}{}
	  +\sum\limits_{d=q,g}\int\frac{\diff x_{b,i}^\prime}{x_{b,i}^\prime}\;
	  P_{bd}(x_{b,i}^\prime)\,\fPDFt{d}{\tfrac{x_{b,i}}{x_{b,i}^\prime}}{\cFtPS\tvar_i}
	\right)	\alphaSt^{n+j}(\muRtsq)\,\BMEp_j(\Phi_j)\,.\hspace*{-12mm}
  \end{split}
\end{equation}
In addition to the transformation properties of the $\BbarME$-function 
of eq.\ \eqref{eq:Bbar-function}, supplemented with the PDF ratios already 
encountered in the leading-order case, additional terms appear. They subtract 
the $\order{\alphaS}$ expansion of these ratios, in order to retain the NLO
accuracy of the merged calculation. Again, please note $\tvar_0=\muFcoresq$.

The same does not hold, however, for the $\mathds{H}$-events. Their 
constituent real-emission matrix elements are defined on $\Phi_{j+1}$ ,
while each subtraction term $\DAMEi{k}$ has its own projection on a 
phase-space point $\Phi_j^k$. Thus, 
\begin{equation}
  \begin{split}
    \lefteqn{\hspace*{-5mm}\HAMEimerge{j}(\Phi_{j+1},\Qcut)}\\
    \,\equiv&\;\RMEmerge_j(\Phi_{j+1};\alphaS,\fPDFs;\muRcore,\muFcore,\cRPS,\cFPS;
                           \{a_i,b_i, x_{a,i},x_{b,i}, \tvar_i\})\;
               \Theta(\Qvar_j-\Qcut)\\
            &{}
               -\sum_k\DAMEimerge{k,j}(\Phi_j^k\cdot\Phi_1^k;
                                       \alphaS,\fPDFs;\muRcorei{k},\muFcorei{k},
                                       \cRPSi{k},\cFPSi{k};\\[-3mm]
            &\hspace*{38mm}
                                       \{a_{i,k},b_{i,k}, x_{a,i,k},x_{b,i,k}, \tvar_{i,k}\})\\
            &\hspace*{12mm}
                \times\;\Theta(\Qvar_j^k-\Qcut)
              \,,
  \end{split}\hspace*{-10mm}
\end{equation}
wherein both $\RMEmerge_j$ and the $\DAMEimerge{k,j}$ separately 
transform as the leading-order counterpart $\BMEmerge_j$. While the 
measure $\Qvar$ on $\Phi_{j+1}$ of $\RMEmerge_j$ is defined to act on the 
underlying $\Phi_j$ after the first cluster step where the real emission 
configuration has been reduced to a Born configuration, it is defined 
directly on each $\Phi_j^k$ in each $\DAMEi{k}$. Infrared 
safety is guaranteed through the infrared safety of their phase-space 
maps, the clustering algorithm and the measure $\Qvar$.

Finally, we consider merging additional multiplicities up to $\jmax$ described
through leading-order matrix elements on top of a next-to-leading order merged 
calculation with up to $\jmaxnlo$ jets, where $\jmax > \jmaxnlo$. The method of 
choice was outlined in
\cite{Hamilton:2010wh,Hoche:2010kg,Hoeche:2012yf,Gehrmann:2012yg,Hoeche:2014rya}
and is historically referred to as 
\MENLOPS. Its methodology is defined as 
\begin{equation}\label{eq:MENLOPS}
  \begin{split}
    \lefteqn{\hspace*{-5mm}\obs^\text{\MEPSatNLO{}+\MENLOPS}}\\
    \,=&\;\sum_{\jmul=0}^\jmaxnlo
          \left[
                \int\diff\Phi_\jmul\;
                \BbarMEmerge_\jmul(\Phi_{\jmul})\;\PStvnlops(O,\Phi_\jmul)
          \right.\\
       &  \left.\hspace*{10mm}{}
                +\int\diff\Phi_{\jmul+1}\;
                 \HAMEimerge{\jmul}(\Phi_{\jmul+1})\;\PStv(O,\Phi_{\jmul+1})
          \right] \\
       &{}+
          \sum_{\jmul=\jmaxnlo+1}^\jmax
          \int\diff\Phi_\jmul\;k_\jmaxnlo(\Phi_{\jmaxnlo+1}(\Phi_\jmul))\;
          \BMEmerge_\jmul(\Phi_\jmul)\;\PStv(O,\Phi_\jmul) \\
   \,=&\;\lim\limits_{N\to\infty}\,\frac{1}{\Ntrial}
          \left\{
                  \sum\limits_{i=1}^{N_\mathds{S}}
                  \sum\limits_{j=0}^{j_\text{max}^\text{\textsc{nlo}}}\;
                  \BbarMEmerge_j(\Phi_{j,i})\;\PStvnlops(O,\Phi_{j,i})
          \right.\\
        & \left.\hspace*{23mm}{}
                 +\sum\limits_{i=1}^{N_\mathds{H}}
                  \sum\limits_{j=0}^{j_\text{max}^\text{\textsc{nlo}}}\;
                  \HAMEimerge{j}(\Phi_{j+1,i})\;\PStv(O,\Phi_{j+1,i})
          \right.\\
        & \left.\hspace*{23mm}{}
                 +\sum_{i=1}^{N_\text{LO}}
                  \sum_{\jmul=\jmaxnlo+1}^{\jmax}\;
                  k_\jmaxnlo(\Phi_{\jmaxnlo+1}(\Phi_\jmul))\;
                  \BMEmerge_j(\Phi_{j,i})\;\PStv(O,\Phi_{j,i})
          \right\}\,,\hspace*{-15mm}
  \end{split}
\end{equation}
with $N=N_\mathds{S}+N_\mathds{H}+N_\text{LO}$. A differential $K$-factor
$k_\jmaxnlo$ is a applied to the higher-multiplicity leading-order 
matrix elements in order to facilitate a smooth transition across $\Qcut$. 
It has the form 
\begin{equation}
  \begin{split}\label{eq:loc-k-fac}
  k_m(\Phi_{m+1})
  \,=&\;
         \frac{\BbarME_m(\Phi_m)}{\BME_m(\Phi_m)}
         \left(
               1-\frac{\HAMEi{m}(\Phi_{m+1})}{\RME_m(\Phi_{m+1})}
         \right)
         +\frac{\HAMEi{m}(\Phi_{m+1})}{\RME_m(\Phi_{m+1})}\;,
  \end{split}
\end{equation}
and therefore moulds the $\BME_{\jmaxnlo+1}$ into the same form as the 
$\HAMEi{\jmaxnlo}$ it is replacing. The projection 
$\Phi_{\jmaxnlo+1}(\Phi_\jmul)$ for $\jmul>\jmaxnlo+1$ is defined through 
its cluster history, as is $\Phi_m(\Phi_{m+1})$ inside $k_m$ itself. 
When now changing the parameters of the calculation, $\alphaS\to\alphaSt$, 
$\fPDFs\to\fPDFst$, $\muR\to\muRt$ and $\muF\to\muFt$, $k_\jmaxnlo$ transforms 
as a composite object in terms of its constituents, cf.\ Sec.\ \ref{sec:fo} 
and \ref{sec:parton_shower}. The scales are set directly by the 
$\BMEmerge_j$ process. Of course, in the interest of decreased computational 
costs one may decide to choose $k_m\equiv 1$ throughout at the cost of 
larger merging systematics. Similarly, if an electroweak cluster history 
leads to a changed signature in $\Phi_{m+1}$, $k_m\equiv 1$ is chosen.

\subsection{Validation}

\begin{figure}
  \centering
  \includegraphics[]{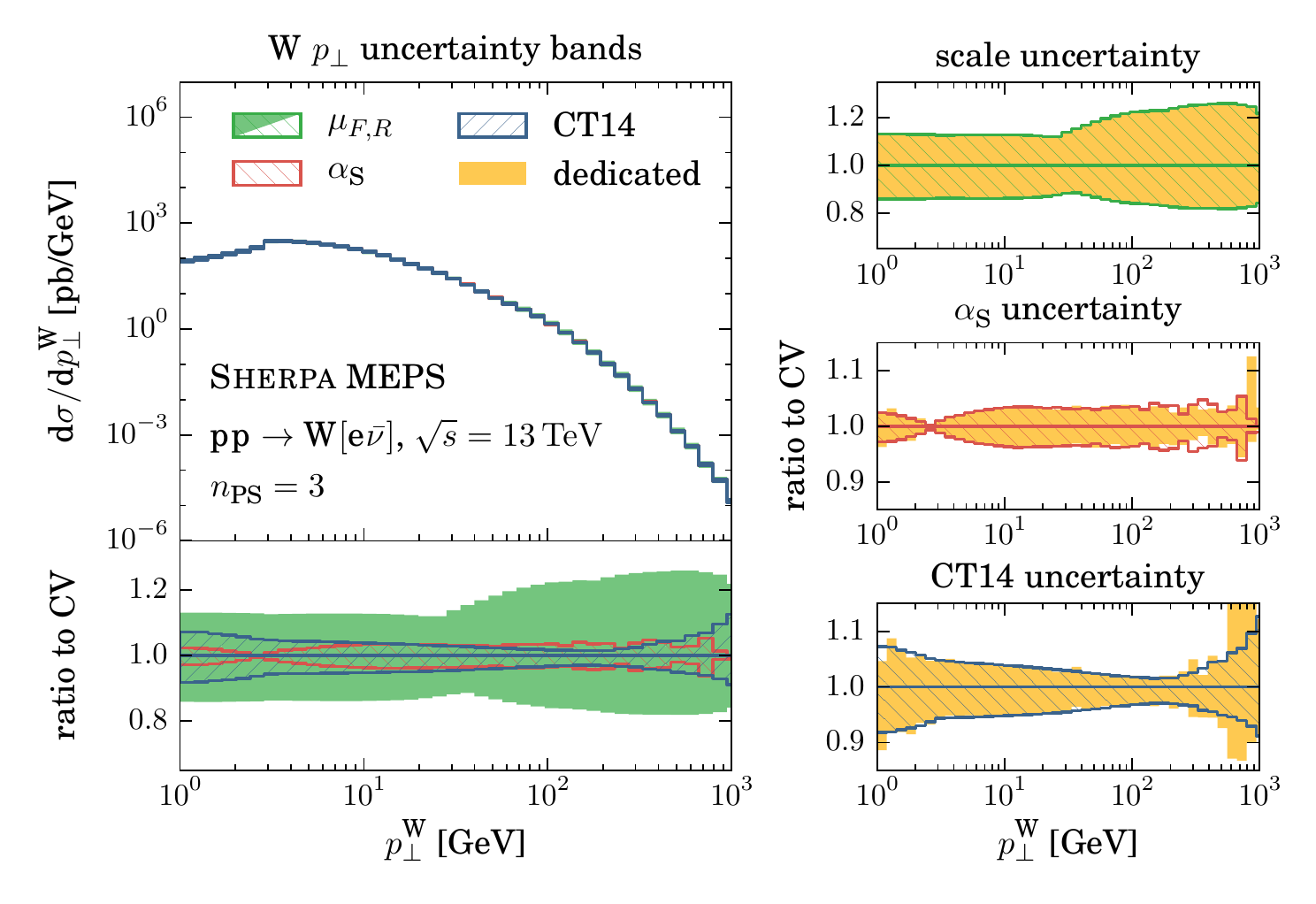}
  \includegraphics[]{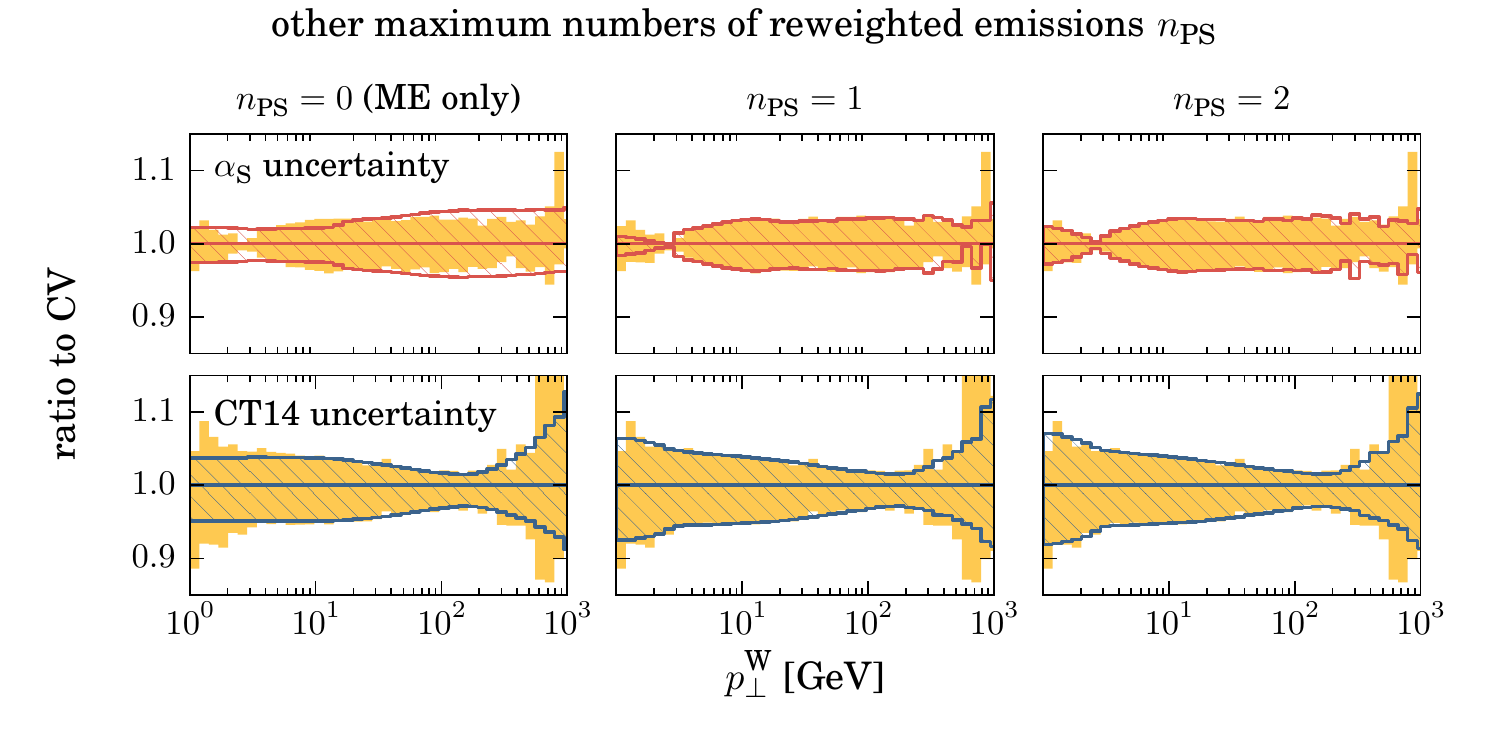}
  \caption{The same as in Figs.~\ref{fig:lops-w-pt} and ~\ref{fig:nlops-w-pt},
      but for a multijet-merged generation with \LO matrix elements for 0-, 1-
      and 2-jet multiplicities.  The uncertainty bands are calculated by
      reweighting the ME and a maximum number of emissions $n_\text{PS}$ of PS emissions.
      In the upper four plots, $n_\text{PS}=3$, thus up to three emissions are
      reweighted.  In the lower plots, $n_\text{PS}$ is varied for comparison.  Again, we
      find a saturation when reproducing dedicated calculations for $n_\text{PS} \geq 2$,
      with no further improvement when $n_\text{PS}$ is increased from 2 to 3.}
  \label{fig:meps-w-pt}
\end{figure}

\begin{figure}
  \centering
  \includegraphics[]{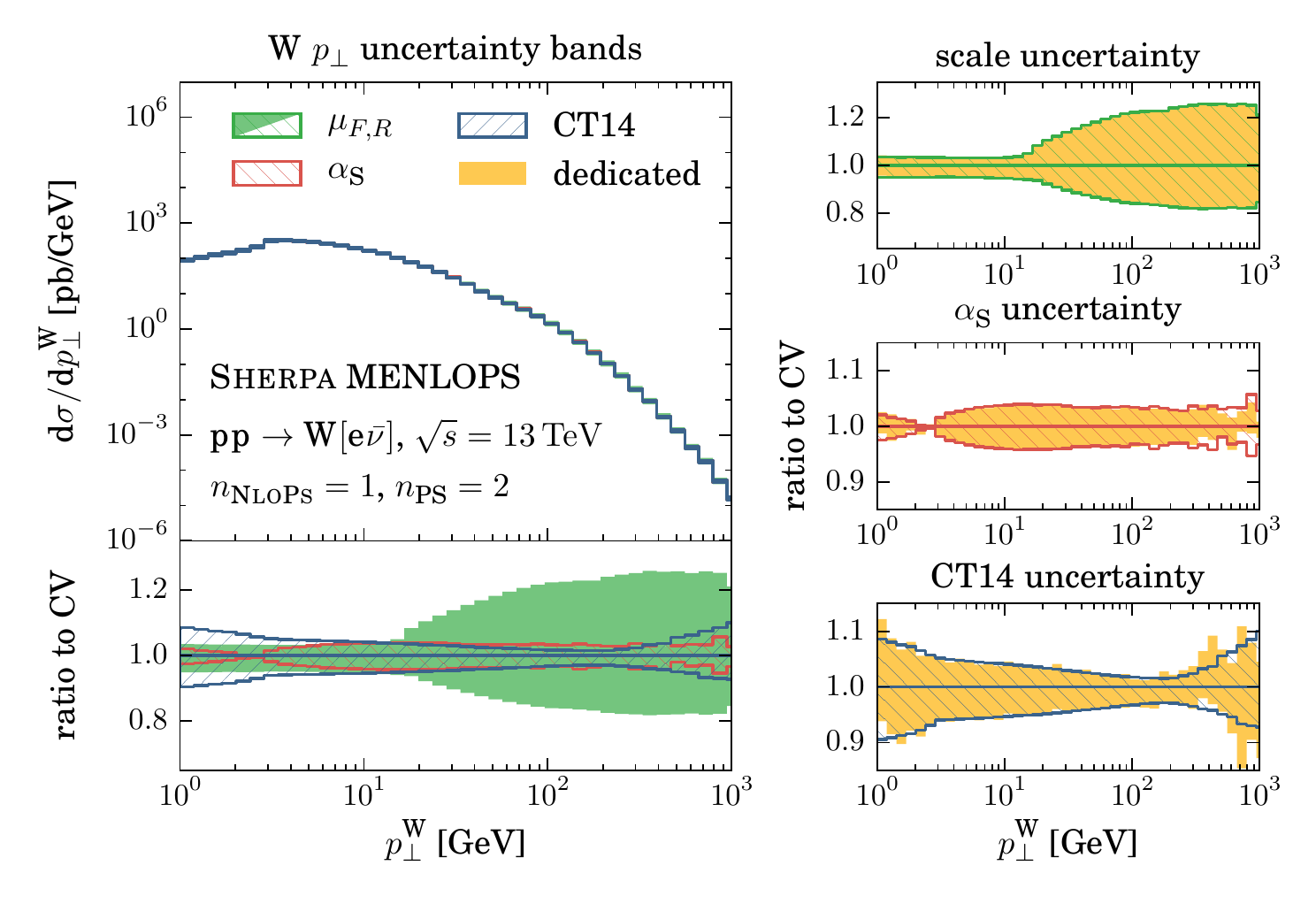}
  \includegraphics[]{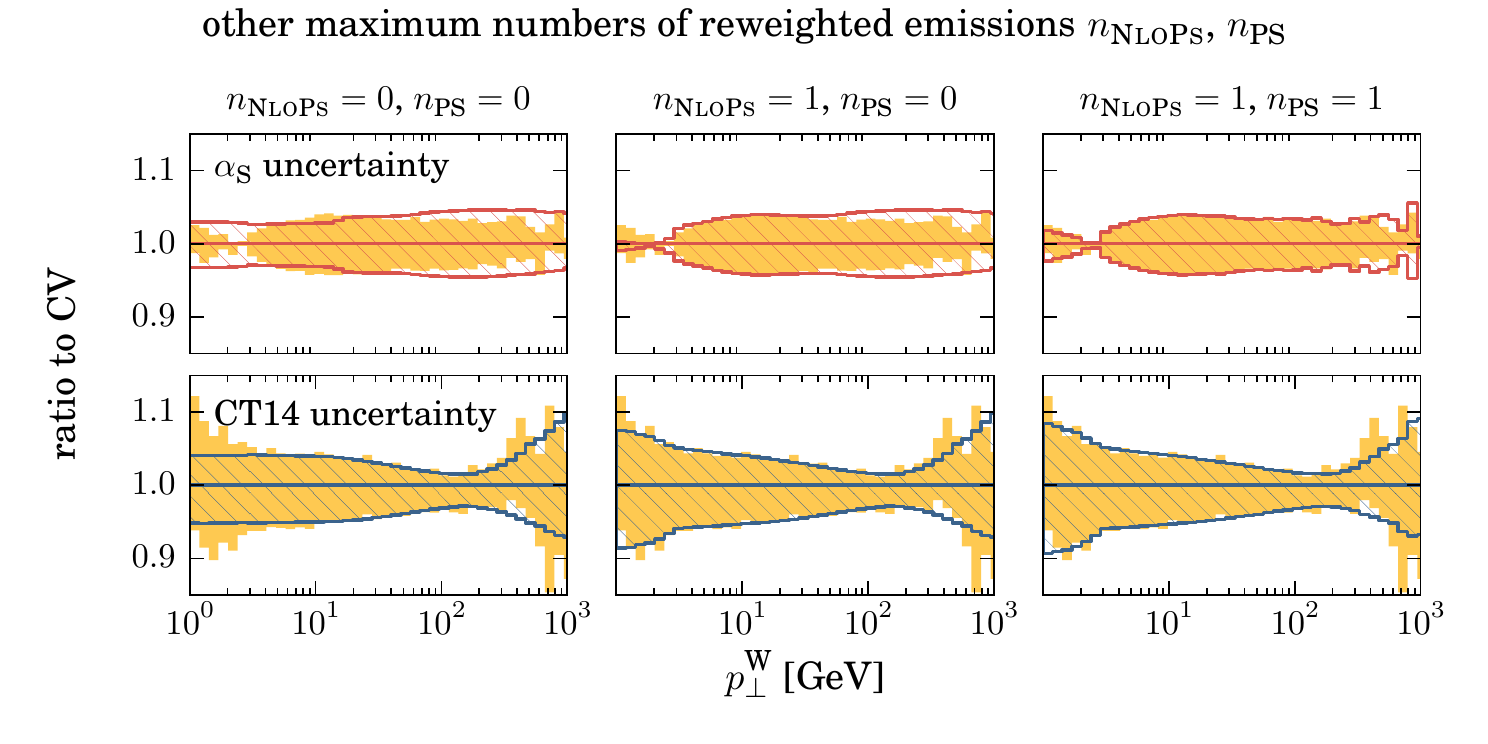}
  \caption{The same as in Figs.~\ref{fig:lops-w-pt}, \ref{fig:nlops-w-pt}
      and~\ref{fig:meps-w-pt}, but for a multijet-merged generation with one
      \NLO matrix element for the 0-jet multiplicity, and \LO matrix
      elements for the 1- and 2-jet multiplicities.  The uncertainty bands are
      calculated by reweighting the ME and a maximum number of emissions from
      the \MCatNLO ($n_\NLOPS$) and the ordinary PS ($n_\text{PS}$).  In the
      upper four plots, $n_\NLOPS=1$ and $n_\text{PS}=2$, thus up to three
      emissions are reweighted.  In the lower plots, both $n$ are varied for
      comparison.  Again, we find a saturation when reproducing dedicated
      calculations for $n_\NLOPS + n_\text{PS} \geq 2$, with no further
      improvement when $n_\text{PS}$ is increased from 1 to 2.}
  \label{fig:menlops-w-pt}
\end{figure}

\begin{figure}
  \centering
  \includegraphics[]{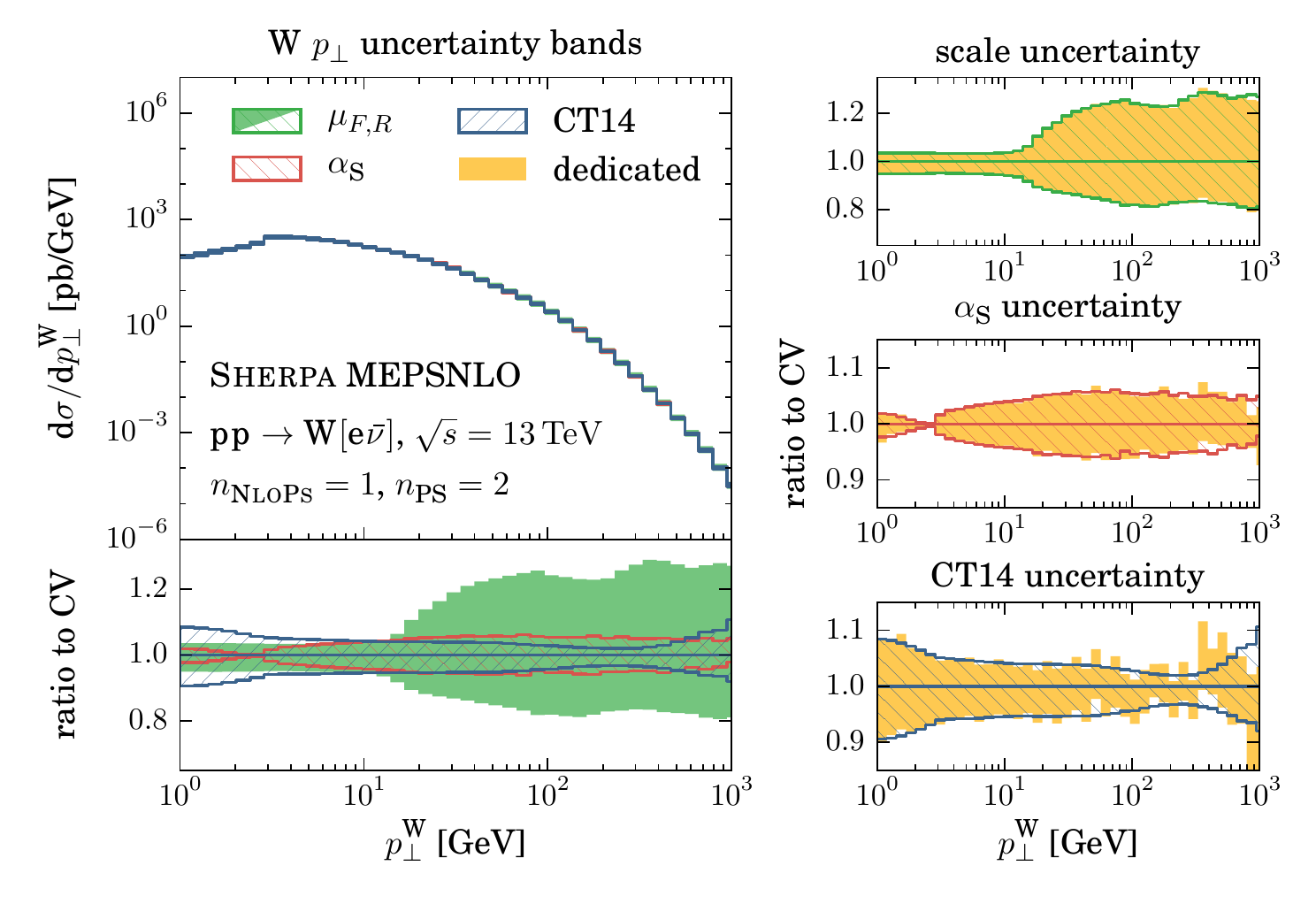}
  \includegraphics[]{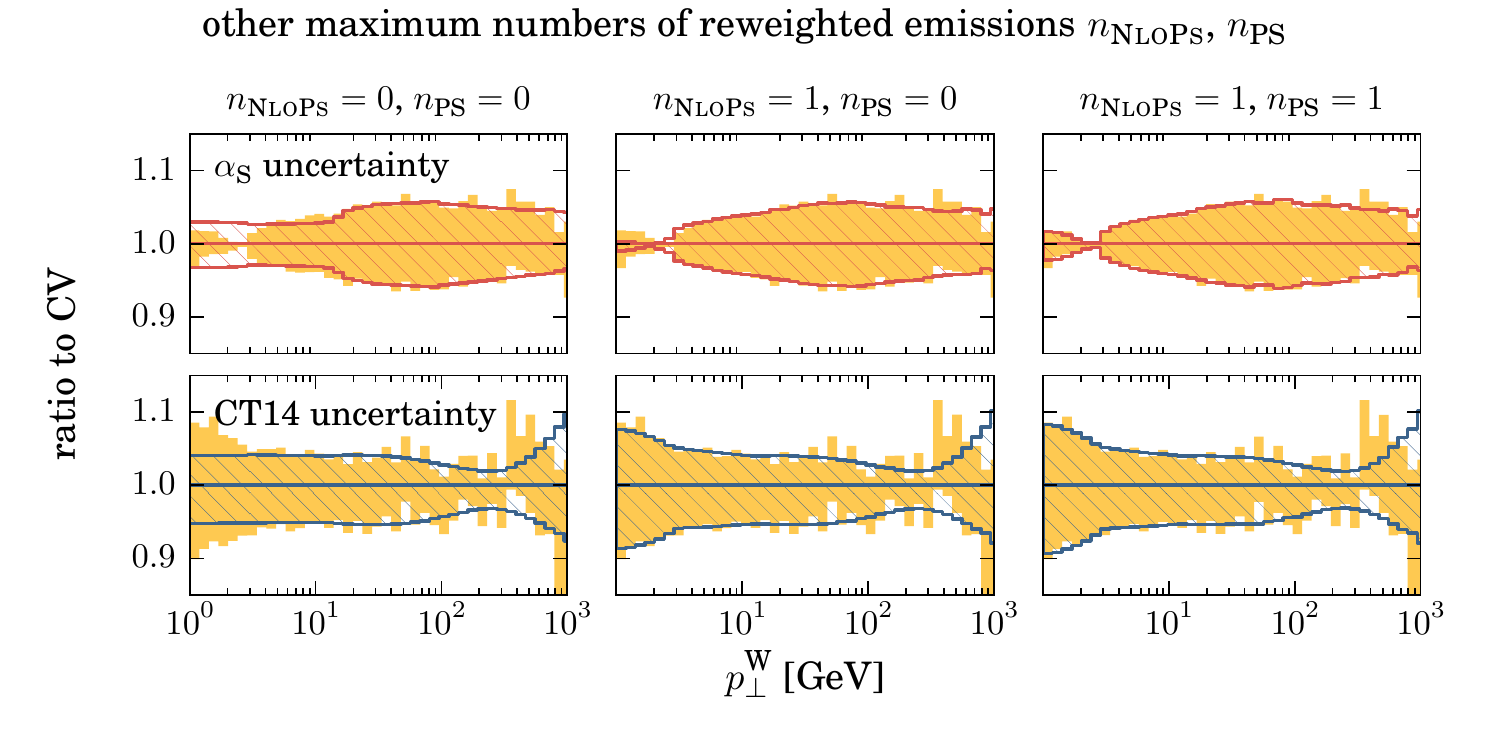}
  \caption{The same as in Figs.~\ref{fig:lops-w-pt}, \ref{fig:nlops-w-pt},
      \ref{fig:meps-w-pt} and~\ref{fig:menlops-w-pt}, but for a multijet
      merged generation with \NLO matrix elements for the 0- and 1-jet
      multiplicities.  The uncertainty bands are calculated by reweighting the
      ME and a maximum number of emissions from the \MCatNLO ($n_\NLOPS$) and
      the ordinary PS ($n_\text{PS}$).  In the upper four plots, $n_\NLOPS=1$
      and $=n_\text{PS}=2$, thus up to three emissions are reweighted.  In the
      lower plots, both $n$ are varied for comparison.  Again, we find a
      saturation when reproducing dedicated calculations for $n_\NLOPS +
      n_\text{PS} \geq 2$, with no further improvement when $n_\text{PS}$ is
      increased from 1 to 2.}
  \label{fig:mepsnlo-w-pt}
\end{figure}

The reweighting for multijet-merged calculations as discussed in the previous
sections has been implemented within \Sherpa with the \CSShower for
leading-order matrix elements (\MEPSatLO), next-to-leading order matrix
elements (\MEPSatNLO) and next-to-leading-order matrix elements with additional
leading-order ones on top (\MENLOPS). For the validation, we again perform
closure tests between reweighted and dedicated predictions for the transverse
momentum of the \PW boson in Figs.~\ref{fig:meps-w-pt}, \ref{fig:menlops-w-pt}
and \ref{fig:mepsnlo-w-pt}. As before, the uncertainty bands are the ones
defined in Table~\ref{tab:variations}, with the CT14 PDF set.  For all merged
calculations, we employ a merging cut of $\Qcut=\SI{20}{GeV}$.

For the \MEPSatLO validation in Fig.~\ref{fig:meps-w-pt}, we combine \LO matrix
elements for 0-, 1- and 2-jet multiplicities, obtained from
\Comix~\cite{Gleisberg:2008fv}.  We can observe that we populate 
a much larger phase space than for a mere \LOPS calculation in terms of 
$p_\perp^{\PW}$. Below the merging cut (i.e.\ $p_\perp^{\PW}
\lesssim \SI{20}{GeV}$), the scale uncertainty band is equal to the one of the
\LOPS calculation. For higher $p_\perp^{\PW}$, the scale uncertainty increases
corresponding to the larger uncertainty of the higher-multiplicity matrix
elements, that contribute renormalisation scale uncertainties.

In Fig.~\ref{fig:menlops-w-pt}, we consider the \MENLOPS case.
We combine an \NLO matrix element for the 0-jet
multiplicities with \LO matrix elements for the 1- and 2-jet multiplicities.
The scale uncertainty for low $p_\perp^{\PW}$ values now features the reduced
scale uncertainty, that we already have seen in the \NLOPS validation.

The same is true in the \MEPSatNLO case depicted in Fig.~\ref{fig:mepsnlo-w-pt}. 
A direct comparison of the scale uncertainties to the \MENLOPS case is not
straightforward though, as we combine \NLO matrix elements for the 0- and
the 1-jet multiplicity, where the virtual amplitudes are obtained from \BlackHat~\cite{Berger:2008sj}. Hence, the 2-jet multiplicity is described at leading 
order through the 1-jet $\mathds{H}$-events. As such, the set-up is not a 
simple upgrade from our \MENLOPS calculation.

In all multijet-merging validations, we find a similar behaviour with respect
to the imprint of including emissions in the reweighting. For $n_{\NLOPS} +
n_\PS = 2$, the dedicated calculations are well reproduced, and no further
improvement is found for $n_{\NLOPS} + n_\PS = 3$.  It is noteworthy, that for
the \MENLOPS case we find a worse reproduction for $n_{\NLOPS} = 1$ and  $n_\PS
= 0$ compared to the \NLOPS and the \MEPSatNLO cases.  This originates in the
fact that in the latter two cases, we enable the reweighting of emissions off
$\mathds{S}$-events at all involved multiplicities, whereas in the \MENLOPS
case only the first of the three multiplicities is affected, because the other
two are at \LO and therefore do not have $\mathds{S}$-events. Thus, the
overall importance of the $\mathds{S}$ emission reweighting gets restricted 
to the region below $\Qcut$ of the 1-jet configuration in the \MENLOPS case.

\section{Consistent variations}

In general, the renormalisation and factorisation scales, $\alpha_s$ and the
PDFs should be varied consistently throughout any of the presented
calculations. While at fixed order the situation is clear, the matched and
merged approaches allow for some degree of freedom regarding partial variations
while still retaining their respective accuracies.

In the simplest case, LOPS, $\mu_R$ and $\mu_F$ of the short distance cross
section and the parton shower may be varied independently as these variations
can be expressed as higher-order terms in a perturbative expansion in the
coupling parameter $\alpha_s$. This is not the case for $\alpha_s$ itself
and the PDFs as they are fixed through measured input values and 
parametrisations. Changes in these input values cannot be expressed 
as simple higher-order terms. Thus they need to be chosen consistently 
throughout. 

Similarly, in NLOPS calculations, the renormalisation and factorisation scales
may be varied in the matrix element ($\overline{\mathrm{B}}$ and
$\mathrm{H}_A$) or the parton shower ($\mathrm{PS}_\text{\textsc{NloPs}}$ and
$\mathrm{PS}$) separately without losing neither the fixed-order nor
the resummation accuracy. As the pseudo-subtraction through the $\mathrm{D}_A$
in any case employs different scales in $\mathrm{PS}_\text{\textsc{NloPs}}$ and
the $\overline{\mathrm{B}}$ and $\mathrm{H}_A$ functions, it always leaves
remainders of $\mathcal{O}(\alpha_s^2)$. Hence, further scale variations in
either one, the short-distance cross sections or the
$\mathrm{PS}_\text{\textsc{NloPs}}$, do not worsen the nominal accuracy of the
method.  Retaining the logarithmic accuracy of the parton shower on the other
hand requires identical renormalisation and factorisation scales throughout all
resummation-relevant components, i.e.\ $\mathrm{PS}_\text{\textsc{NloPs}}$ and
$\mathrm{PS}$.  Again, variations in $\alpha_s$ or the PDFs need to be
consistent throughout the calculation.

The multijet-merged calculations impose further constraints since they treat
multijet matrix elements and parton-shower emissions on the same footing. The
notation of the scales already reflects this for $\mu_R$ and $\mu_F$. In their
definitions only the core scales remain as free parameters and may be varied
independently.  Again, the $\alpha_s$ and PDF parametrisations need to be the
same throughout.

\section{Conclusions}
\label{sec:conclusions}

In this publication we have presented the implementation
and validation of reweighting techniques allowing for the
fast and efficient evaluation of perturbative systematic 
uncertainties in the \Sherpa event-generator framework. 
We have lifted the available techniques for the 
determination of PDF, $\alpha_s$ and scale uncertainties 
in leading- and next-to-leading order QCD calculations to
include the respective variations in parton-shower 
simulations. In turn we provide the means to perform 
consistent uncertainty evaluations for multijet-merged
simulations based on leading- or next-to-leading-order
accurate matrix elements of varying multiplicity matched
with parton showers. The foundation for our reweighting
method is the knowledge of the very dependence structure
of the perturbative calculations on the parameters to be 
varied. For the fixed-order components this amounts to the
corresponding decomposition of the Catani--Seymour dipole
subtraction terms. This needed to be supplemented by the
reweighting of the parametric dependences of the parton
shower treated through the Sudakov Veto Algorithm. 

With our extensive validation we have been able to 
prove on the one-hand-side the correctness of the 
implementation and have, furthermore, been able to 
illustrate the importance of parton-shower reweighting
for reliable uncertainty estimates. With comparably 
little additional computational costs this allows for 
the on-the-fly determination of PDF, $\alpha_S$ and
scale uncertainties based on one single generator run,
that, otherwise, would require explicit re-computations.
The overall reduction in CPU time is by a factor of about
3 to 20, depending on the event-generation mode used, 
see App.~\ref{app:timing}. The variational event weights 
provided are easily accessible through the \HepMC event 
record and are furthermore consistently handed over to 
the \Rivet analysis software by the corresponding \Sherpa 
interface.

The methods presented in this publication are ideally 
suited for event-wise uncertainty estimates
and can readily be used in arbitrary theoretical and
experimental analyses. An extension to 
next-to-next-to-leading-order QCD calculations possibly 
dressed with parton showers, as presented in 
\cite{Hoeche:2014aia,Hoche:2014dla,Dawson:2016ysj}, is
straightforward and planned for the near future. 

The decomposition of the fixed-order part of
QCD calculations employed here is also a necessary 
ingredient to produce cross-section grids as provided 
by the \mbox{\APPLgrid}~\cite{Carli:2010rw} and 
\FastNLO \cite{Kluge:2006xs,Britzger:2012bs} tools. These 
store the perturbative coefficients for a certain observable
calculation discretised in $Q^2$ and $x$. Using
interpolation methods, this allows for the {\em a posteriori} 
inclusion of PDFs, $\alphaS$ and variations of the
renormalisation and factorisation scales. In turn, such 
techniques are well suited for (combined) fits of PDFs 
and $\alphaS$ that require a multitude of re-computations 
of the theoretical predictions. Over the last years, tools
have been developed that automate the projection of arbitrary
next-to-leading-order QCD calculations onto such grids,
namely the \aMCfast \cite{Bertone:2014zva} and the 
\MCgrid \cite{DelDebbio:2013kxa,Bothmann:2015dba,Bothmann:2015woa}
packages. The first one produces \APPLgrid{}s with 
\MGMCatNLO \cite{Alwall:2014hca}, the latter  
\APPLgrid{}s or \FastNLO grids from \Sherpa events projected on
the observables through \Rivet. The \mbox{\APFELgrid} tool 
\cite{Bertone:2016lga} provides an improved convolution method 
for use with \APPLgrid files that furthermore speeds-up the 
re-evaluations. 

However, none of these approaches includes generic
parton-shower effects, i.e. the parametric dependence of 
the shower component on the PDFs and $\alphaS$ is ignored. 
With the methods presented in this publication we are 
confident that we can surmount this limitation and in the 
future provide interpolation grids that properly reflect
shower-resummation effects and allow for the inclusion
of the affected phase-space regions in PDF determinations.

\section*{Acknowledgements}

We acknowledge financial support from BMBF under contract 05H15MGCAA, 
the Swiss National Foundation (SNF) under contract PP00P2–128552 and 
from the EU MCnetITN research network funded under Framework Programme 
7 contract PITN–GA–2012–315877.

\appendix
\section{CPU time measurements}
\label{app:timing}

The benefit of reweighted calculations is given by the saving of CPU time. 
In order to evaluate the gain, we shall compare the event generation time 
of reweighted calculations with the sum of generation times for all 
corresponding dedicated computations. Here we consider both parton-level 
calculations, as well as runs including multiple interactions and 
hadronisation, the typical default in physics analyses applications. 
For the latter it can be expected that the gain in CPU time by using the
reweighting approach is most considerable, as the CPU intense non-perturbative
event generation phases do not need to be re-evaluated. In what follows 
we compare actual event-generation times, neglecting the set-up times of 
the individual runs.%
\footnote{
    If NLO matrix elements at higher multiplicities are needed for an
    event generation, the time needed for the integrator optimisation and the
    process selection weight optimisation can
    be quite substantial, e.g.\ a couple of days. In the case of unweighted
    event generation, this even has to be re-done for every single parameter
    variation, as the channel weights are used for the unweighting. When
    reweighting is used, this is not necessary and so even more CPU time is
    saved.%
}

\begin{figure}
  \centering
  \includegraphics[width=\linewidth]{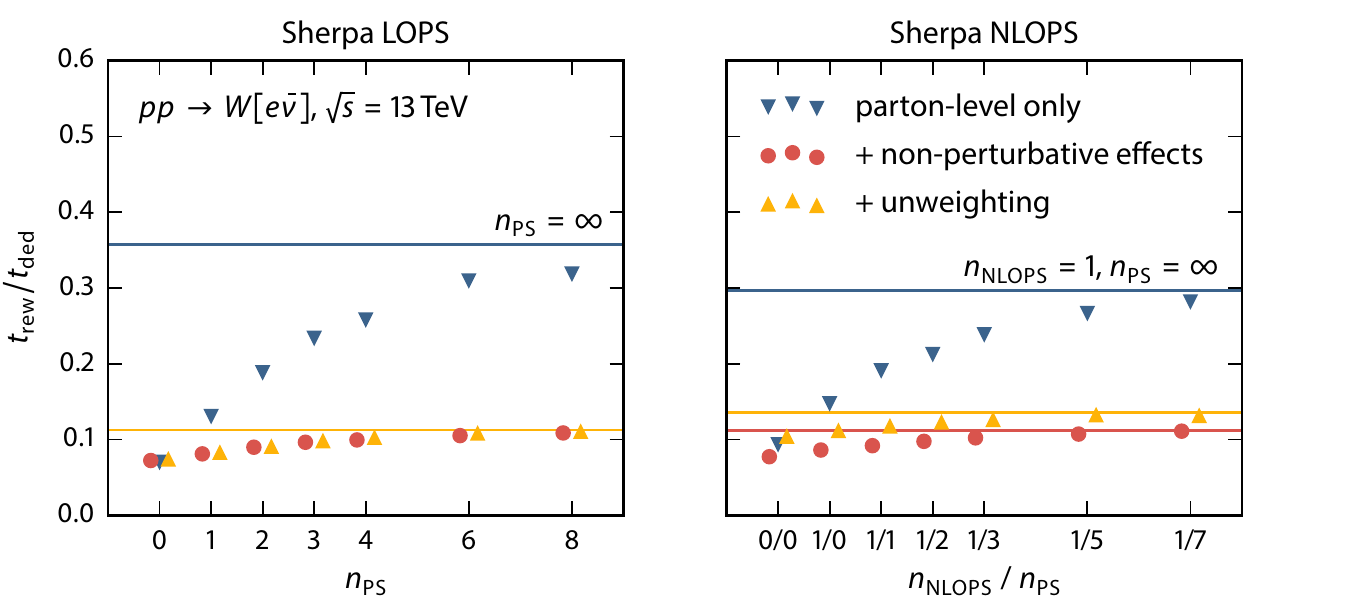}\\[1em]
  \includegraphics[width=\linewidth]{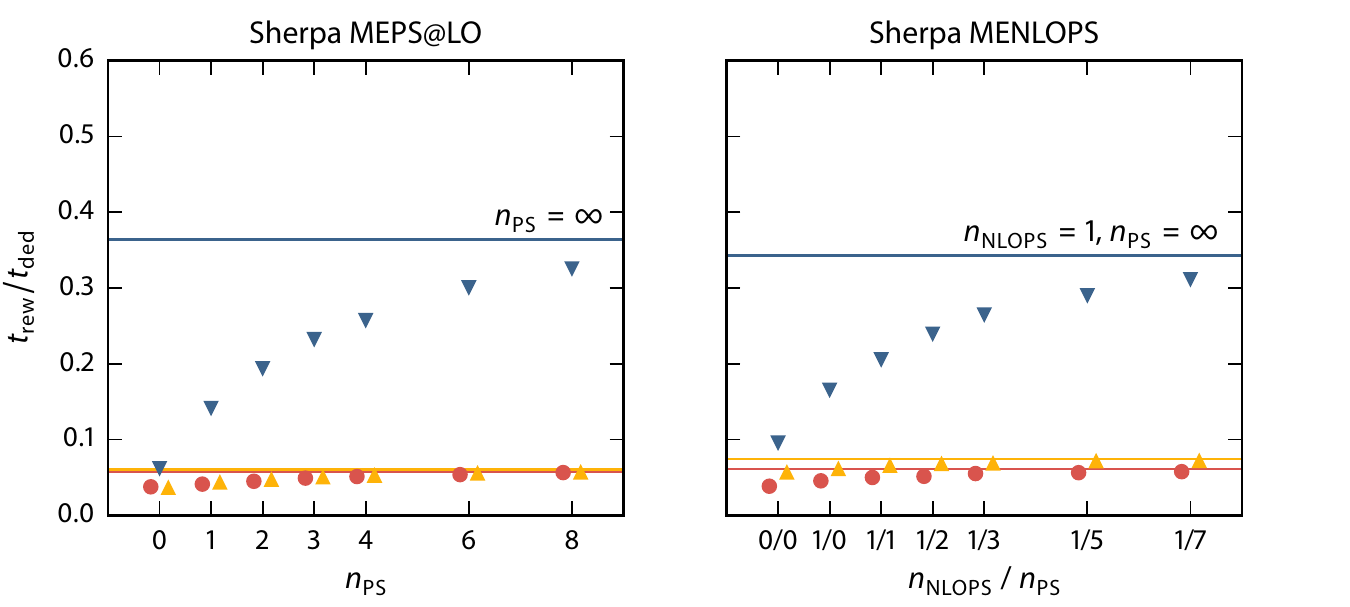}
  \caption{The ratio of CPU time needed for a reweighting
      event generation~$t_\text{rew}$ over the time needed for generating
      predictions for all variations with dedicated runs~$t_\text{ded}$. The
      reweighting includes up to $n_\text{PS} + n_\NLOPS$ parton-shower
      emissions. The sampled variations are listed in
      Tab.~\ref{tab:variations}.
      Parton-level-only results are compared to results for calculations
      including multiple interactions and hadronisation effects (``+
      non-perturbative''), and to calculations where in addition to adding
      non-perturbative effects the events have also been unweighted (``+
      unweighting''). The ratios for reweighting all emissions are indicated
      with a horizontal line.%
      \label{fig:cpu_times}}
\end{figure}

In Fig.~\ref{fig:cpu_times}, we consider event generations using \LOPS, \NLOPS,
\MEPSatLO and \MEPSatNLO calculations for $\HepProcess{\Pproton\Pproton \to
\PW[\Pelectron\APneutrino]}$ at \SI{13}{\TeV}.  The ratio of CPU time between
the reweighting and the dedicated generations is shown for different maximum
numbers of reweighted shower emissions $n_\text{PS} + n_\NLOPS$.  Whether
non-perturbative effects are included or not, the time needed for the
reweighting calculation is below \SI{10}{\percent} of the time needed for
dedicated calculations if only the matrix element is reweighted ($n_\text{PS} =
n_\NLOPS = 0$). The ratio then increases for larger numbers of reweighted
emissions, as their reweighting needs additional time, asymptotically
approaching the value when all parton-shower emissions are reweighted.
For parton-level-only calculations, this ratio is around $0.35$ for \LOPS 
events, and around $0.3$ for \NLOPS events. This reduction can be explained 
due to relatively smaller computational cost of the parton shower as a whole 
when the rest of the calculation is more complex.  Also note that $n_\text{PS}$ for
\LOPS is only equivalent to $n_\text{PS} + n_\NLOPS$ for $\mathcal{S}$ events.
$\mathcal{H}$ events do not feature the \SMCatNLO-emission, and hence for them
$n_\NLOPS$ does not contribute to their reweighting.

For the same reason, when non-perturbative effects are included, that ratio
improves to about $0.1$: The parton shower (and its reweighting) component
plays a relatively smaller r\^{o}le in terms of CPU cycles, when multiple 
interactions and hadronisation are enabled.

If on top of the non-perturbative effects the events are also unweighted, the
ratio does not change in the \LOPS case, but in the \NLOPS case (by about
\SI{20}{\percent}). A reason might be, that only for \NLOPS a sizeable number
of events gets rejected.  For these, the jet evolution and non-perturbative
phases are not performed at all, whereas the matrix-element calculation (and
its reweighting) is always done, for accepted and rejected events alike. The
same is true for the \SMCatNLO{} emission from $\mathds{S}$ events.  As a
consequence, the \emph{relative} cost of the reweighting grows slightly.  A
future improvement of the implementation would postpone these futile
reweightings in unweighted calculations to a time point after the possible
rejection. This of course requires that the dependence of the rejection
probability is negligible. For the observables studied so far this was found to
be true, at least to $\order{10^{-4}}$.

Note that the effective gains will be lower than the results presented in this
section, when we take into account the reduced statistical accuracy which comes
with the parton-shower reweighting. This requires more events to be generated
in a reweighting calculation to reach the same statistical accuracy as in a
dedicated calculation.

\section{Configuring and accessing event-weight variations}
\label{app:keywords}

\Sherpa provides a list of pre-calculated alternative event weights, which are
automatically output to the \HepMC event record \cite{Dobbs:2001ck} 
or directly to an interfaced \Rivet analysis \cite{Buckley:2010ar}. 
For versions of \Sherpa later than v.2.2.0, the variations to
calculate can be specified with the following line
in the \texttt{(run)} section of the \Sherpa run card:
\begin{verbatim}
  VARIATIONS muR2fac1,muF2fac1,PDF1 muR2fac2,muF2fac2,PDF2 ...;
\end{verbatim}
Each variation is characterised by up to three arguments
\begin{description}
   \item[\texttt{muR2fac}]
     a prefactor multiplying the nominal (squared) renormalisation scale
   \item[\texttt{muF2fac}]
     a prefactor multiplying the nominal (squared) factorisation scale
   \item[\texttt{PDF}]
     a parton density and its accompanying $\alphaS$ parametrisation.
\end{description}
This syntax works for all employed scale setters of \Sherpa and both \Sherpa's
internal PDFs and PDFs interfaced through \textsc{Lhapdf5/6} 
\cite{Whalley:2005nh,Buckley:2014ana}.  If trailing arguments are omitted 
from a variation, their default values are used, which is 1.0 for scale factors 
and the PDF set used by \Sherpa for the nominal calculation.

In \HepMC event records (v.\ 2.06 or later), the alternate weights can be accessed as named weights
within the \texttt{HepMC::WeightContainer} of each event. The keys are given in
one of the following formats:
\begin{verbatim}
    MUR<muR2fac>_MUF<muF2fac>_PDF<ID>
    MUR<muR2fac>_MUF<muF2fac>_PDF<ID>_PSMUR<muR2fac>_PSMUF<muF2fac>
\end{verbatim}
The parts in angle brackets are replaced with the respective scale factors and
\textsc{Lhapdf} IDs.  The second form is used, if a factor is applied to the
renormalisation/factori\-sation scale of parton-shower splittings. This includes
splittings within cluster histories determined by the multijet merging
procedure, as discussed in section~\ref{sec:multijet_merging}.  If the scale
reweighting with parton-shower splittings has been enabled (we discuss below
how to do so), the scale factors for \texttt{MUR}, \texttt{MUF} and
\texttt{PSMUR}, \texttt{PSMUF} are always equal, respectively, in the current
implementation.

If the internal \Rivet interface of \Sherpa is used to analyse events
during the generation, one histogram file per variation is written to disk,
along with the nominal one. The file names follow a pattern resembling
the \HepMC weight-container keys as specified above.

\subsection*{Scale variations}

The scale argument can also be specified by enclosing it in square brackets:
\texttt{[mu2fac]}. This syntactic sugar implies both the given factor, its
inverse and the default value.  For example, \texttt{1.0,[4.0]} is equivalent
to \texttt{1.0,4.0 1.0,0.25 1.0,1.0} and therefore triggers up and down
variations of the factorisation scale, along with the central value.  If both
scale factors are enclosed in brackets, they are expanded individually, keeping
the other at its default value of 1.0: Hence, \texttt{[4.0],[4.0]} is
equivalent to the 5-point scale variation \texttt{4.0,1.0 0.25,1.0 1.0,4.0
1.0,0.25 1.0,1.0}.  To include simultaneous variations in the same direction,
both factors can be surrounded by a single pair of brackets.  Thus,
\texttt{[4.0,4.0]} is equivalent to the 7-point scale variation \texttt{4.0,1.0
0.25,1.0 1.0,4.0 1.0,0.25 4.0,4.0 0.25,0.25 1.0,1.0}.

\subsection*{PDF and $\alphaS$ variations}

PDF and $\alphaS$ variations both work by specifying a PDF set through the
\texttt{PDF} argument of a variation. This is because \Sherpa per default uses
the value for $\alphaS(m^2_Z)$ given by the PDF set in use. Therefore an
$\alphaS$ variation can be achieved by using PDF fits for different values of
$\alphaS(m^2_Z)$.

To specify a specific member of a PDF set, its number is given as an additional
argument separated by a slash. Thus, \texttt{1.0,1.0,CT14nlo/38} asks for the
38th member of the \texttt{CT14nlo} PDF set, without modifying the
renormalisation and factorisation scales. If the slash and the number are not
given, the central PDF member is used, i.e. \texttt{CT14nlo} is equivalent to
\texttt{CT14nlo/0}.

\Sherpa can also be asked to do variations for all members of a PDF set by
enclosing it in square brackets. Hence, \texttt{1.0,1.0,[CT14nlo]} is
equivalent to
\begin{verbatim}
  1.0,1.0,CT14nlo/0 1.0,1.0,CT14nlo/1 ... 1.0,1.0,CT14nlo/56
\end{verbatim}
This \texttt{[PDF]}-notation only works with PDFs interfaced through
\textsc{Lhapdf6} \cite{Buckley:2014ana}.  It can be combined with scale factors
that are enclosed in square brackets.  Again, the expansions are done
individually, keeping other arguments at their default values.  This means that
for example  \texttt{1.0,[4.0],[CT14nlo]} is equivalent to \texttt{1.0,[4.0]
1.0,1.0,[CT14nlo]}. Hence, a 7-point scale variation and a full
\texttt{CT14nlo} PDF variation can be requested by
\begin{verbatim}
  VARIATIONS [4.0,4.0],[CT14nlo];
\end{verbatim}

\subsection*{Configuring how variations are calculated}
The following options always affect all variations that are specified by
arguments to the \texttt{VARIATIONS} keyword.
\begin{description}
    \item[\texttt{REWEIGHT\_SPLITTING\_ALPHAS\_SCALES}] (default: 0) If this is set to
        1, the renormalisation scale factor is applied to the $\alphaS$
        argument of individual splittings, instead of applying it only to the
        overall renormalisation scale, see section~\ref{sec:merging-lo}.  This
        means that parton-shower splittings are only included in the rescaling,
        if this option is enabled. In the notation of
        sections~\ref{sec:parton_shower} and~\ref{sec:multijet_merging}, this
        sets $\cRtPS = \muRt / \muR$.
    \item[\texttt{REWEIGHT\_SPLITTING\_PDF\_SCALES}] (default: 0) If this is set to
        1, the factorisation scale factor is also applied to PDF scale arguments within
        shower splittings (and intermediate cluster history PDF ratios), and
        not only to the core-process PDFs. In the notation of
        sections~\ref{sec:parton_shower} and~\ref{sec:multijet_merging}, this
        sets $\cFtPS = \muFt / \muF$.
    \item[\texttt{REWEIGHT\_MAXEM}] (default: -1) This option specifies the
        number of ordinary parton-shower emissions included in the reweighting
        per event. If this is set to 0, no emission is reweighted.  The default
        value -1 means that all emissions should be reweighted.
    \item[\texttt{REWEIGHT\_MCATNLO\_EM}] (default: 1) If this is set to 0, the
        single parton-shower emission within the \MCatNLO contribution is not 
        reweighted.
    \item[\texttt{VARIATIONS\_INCLUDE\_CV}] (default: 1) If this is set to 0,
        the behaviour of the square bracket syntax is changed, such that the
        central-value variation is not included when expanding a parameter in
        square brackets. It is recommended not to disable it, such that one can
        do a closure test between the dedicated calculation and the
        reweighting. However, in CPU intensive applications, this setting can
        be used to omit this one obsolete variation while still making use of
        the convenient square-bracket syntax.
\end{description}

\clearpage

\bibliographystyle{JHEP}
\bibliography{main}

\end{document}